\documentclass[12pt,preprint]{aastex}
\usepackage{psfig}

\doublespace
\newcommand{\Ha}{H$\alpha$}

\newcommand{\GH}{HI 1225$+$01}
\newcommand{\Lya}{Ly$\alpha$}

\newcommand{\Lyf}{Ly$\alpha$ forest}

\newcommand{\NII}{[\ion{N}{2}]}
\newcommand{\surfbright}{ergs~cm$^{-2}$~s$^{-1}$~arcsec$^{-2}$}
\newcommand{\surfperA}{ergs~cm$^{-2}$~s$^{-1}$~arcsec$^{-2}$~\AA$^{-1}$}
\newcommand{\cmsq}{cm$^{-2}$}
\newcommand{\kms}{km~s$^{-1}$}

\newcommand{\Msun}{\ifmmode M_{\odot} \else $M_{\odot}$\fi}

\shorttitle{Limit to Ionizing Background Radiation}
\shortauthors{Weymann et al.}

\begin{document}

\title{A New Observational Upper Limit to the Low Redshift Ionizing
Background Radiation}

\author{Ray J. Weymann}

\affil{Carnegie Observatories, 813 Santa Barbara Street, Pasadena CA 91101}

\email{rjw@ociw.edu}

\author{Stuart N. Vogel}

\affil{Department of Astronomy, University of Maryland, College Park, MD 20742}

\email{vogel@astro.umd.edu}

\author{Sylvain Veilleux}

\affil{Department of Astronomy, University of Maryland, College Park, MD 20742}

\email{veilleux@astro.umd.edu}

\author{Harland W. Epps}

\affil{Lick Observatory, University of California Santa Cruz, Santa
Cruz, CA 95064}

\email{epps@ucolick.org}

\begin{abstract}

We report a new Fabry-Perot search for \Ha\ emission from the
intergalactic cloud \GH\ in an attempt to measure the low redshift
ionizing background radiation. We set a new $2\sigma$ upper limit on
\Ha\ emission of 8 mR ($5\times 10^{-20}$ \surfbright).
Conversion of this limit to limits on the strength of the ionizing background
requires knowledge of the ratio of the projected to total surface area
of this cloud, which is uncertain. We discuss the plausible range of
this ratio, and within this range find that the strength of the ionizing
backround is in the lower range of, but consistent with, previous observational
and theoretical estimates.
%Adopting the recent estimate for the inclination of \GH,
%this upper limit implies a strength for the ionizing background which
%is compatible with estimates based upon simulations of the \Lyf, but
%below the recent best estimates for the combined effects of AGNs and
%hot stars in star-forming galaxies.

\end{abstract}

\section{INTRODUCTION}

The value of the local ionizing flux is of interest for several
reasons: 1) It plays an important role in setting the level of
ionization of the low density hydrogen gas which gives rise to the
\Lya\ forest. Recent simulations have been remarkably successful in
reproducing the main features of the \Lya\ forest from high redshifts
to zero redshift \citep{dave}.  As discussed in detail in \S4, for a
given set of cosmological parameters, these simulations are now able
to make predictions about the value of the local H~I photoionization
rate, and it would be valuable to determine empirically the value of
this rate to compare it with the simulations. With a sufficiently
accurate determination of its value and further forthcoming data on
the low redshift \Lya\ forest, together with further improvements in
the simulations it may be possible to set some constraints on these
cosmological parameters, especially the baryon number.  2) A limit on
the photoionization rate in locations which are free of local ionizing
sources (i.e. star-forming regions or ionization due to high energy
particles in the neighborhood of galaxies) can then be used to assess
the relative importance of such local sources near a galaxy compared
to the non--local, global value. In particular, there is currently some 
controversy concerning the ionization source for some high velocity
clouds in our own galaxy \citep{ben,bhput}.  3) Estimates of the
photoionization rate due to the cumulative effect of active galactic
nuclei can now be carried out with fair confidence \citep[hereafter
SRGPF]{shull}; but there remain significant uncertainties associated
primarily with the luminosity functions of these objects and with the
evaluation of the opacity associated with intervening neutral
hydrogen.  4) Similar remarks apply to the contribution from escaping
radiation shortward of the Lyman limit from hot stars in
galaxies. However, this estimate is also dependent upon the fraction
of the Lyman continuum radiation assumed to escape from the galaxies,
and direct observational evidence about the magnitude of this escape
fraction is still somewhat scanty.  

For these reasons, we have previously set limits on the value of the
non-local ionization rate at zero redshift by setting limits on the
\Ha\ flux from an isolated, large, H~I cloud which is optically thick
to the Lyman continuum \citep[hereafter VWRH]{vwrh}.  The essence of
the method is very simple.  Illumination by ionizing radiation of an
H~I cloud that is optically thick to the Lyman continuum results in
\Ha\ emission that is well described by Case B recombination theory.
Accordingly, approximately one \Ha\ photon escapes for every two
incident Lyman continuum photons; the exact value depends only mildly
on parameters that are reasonably well known.  The one significant
uncertainty is the ratio of the observed projected surface area to
that illuminated by the ionizing radiation; however, the uncertainty
is perhaps a factor of two, and can be better determined for objects
where the geometry is known, such as galactic disks.  The expected
\Ha\ radiation is extremely faint --- based on predictions for the low
redshift metagalactic ionizing flux (e.g., SRGPF), the \Ha\ surface
brightness is expected to be only a few milliRayleighs (for \Ha, 1
mR $=5.67\times 10^{-21}$\,\surfbright).  By comparison, the observed
\Ha\ emission from objects ionized by other sources is significantly
brighter: high velocity clouds known to be within a few kpc are
typically 60--200 mR \citep{tufte}; most compact high velocity clouds
lie between 30 to 300 mR \citep{ben}; the Magellanic Stream ranges up
to more than 1000 mR \citep{ww96}; a brightness of 90 mR was measured
in the outer disk of NGC 253 \citep{bfq}; and the diffuse extraplanar
ionized gas imaged in edge-on spirals is typically brighter than 1000
mR \citep{rfk,vcbh}.  As a result of its expected faintness,
detection of \Ha\ emission due to ionization by the metagalactic
ionization requires observation of extended clouds using specialized
techniques.

There exist other approaches to an empirical value for the global zero
redshift ionization rate.  However, several of these methods suffer from
significant limitations.  At higher redshifts, the proximity effect
\citep{bajtlik,bechtold94} has been used to estimate the value of the
metagalactic ionizing flux; however at low redshift the sparseness of
higher resolution data and the relative sparseness of the \Lyf\
itself makes this method currently less useful.  In addition,
there are some issues about the interpretation of the proximity effect
(e.g., the clustering properties of matter in the vicinity of QSOs; see
\citet{scott} for further discussion).

%The techniques using observations of the outer
%H~I disks of galaxies suffer from ionization by local sources
%\citep{bfq}; in fact, knowledge of the actual value of the
%metagalactic ionizing flux would greatly help in modeling the extent
%and structure of gaseous galactic disks \citep{maloney,dsm94}. 

%RJWREVBEG
%%% 2nd revision starts here and ends at %%%###

A potentially very powerful technique involves observing the rapid transition
from a mostly neutral hydrogen disk to a mostly ionized disk as one moves outward
along the radius of spiral galaxies. From the value of this
transition column density, one can infer the flux of the background ionizing
radiation given a model for the vertical distribution of gas.
The basic idea for this technique goes back to papers by 
\citet{suny} and \citet{felt}.
The historical development of this idea is summarized by 
 \citet{maloney} who applied it to realistic galaxy models.
Maloney gives a careful
treatment of the radiative transfer equation 
and discusses a wide range of
galaxy parameters. His results show that
the derived ionizing flux is relatively insensitive to the dynamical models adopted for the galaxy.
The application to NGC3198, which has the most sensitive HI observations in this context,
and whose geometry and dynamics are not very uncertain, yielded a transition column density of
about $5 \, \times 10^{19} \, cm^{-2}$ and from this Maloney deduced that the flux of ionizing
photons is in the range $ 5 \, \times \, 10^3$ to $  5 \, \times \, 10^4 \, { \rm photons \, cm^{-2}
\, sec^{-1}}$. Shortly
after this work \citet{dsm94} published a similar analysis and confirmed
these results. They also showed that the results are not very sensitive to the assumed spectral
index of the ionizing background. 
%%The principle contributor to the uncertainty in
%%the ionizing flux appears to 
%% be knowledge of the radial variation of the total column density of hydrogen at these large radii. 

Application
of this method to other objects has also been carried out. In particular, as noted above,
measurements of H$\alpha$ emission beyond the HI truncation radius were obtained by \citet{bfq}
in NGC253. In this case, however, these authors noted that the inferred surface brightness
was higher than could likely arise only from the metagalactic radiation field. In addition, the
ratio of  [NII] $\lambda$6548 to H$\alpha$ emission was unusually high. These authors favored a model
in which a warp in the outer disk allowed ionizating flux from the inner young star population
to dominate the ionization. In NGC253 the ratio of the HI disk radius to the optical disk radius
is considerably smaller than in NGC3198. Nevertheless, \citet{dsm94} do not reject the
possibility that hidden local sources of ionization may contribute significantly to the ionizing
flux even in NGC3198. 

%%%Finally, 
We note that there appear to be instances where
no abrupt truncation of H~I occurs to column densities which are 
smaller than $\sim$ $5 \, \times 10^{19} \, {\rm cm^{-2}}$. For example, \citet{wsso} present HI
images of NGC 289, a giant gas-rich low-surface-brightness spiral, and
show HI contours down to a column density of $1 \times 10^{19}$ \cmsq;
they note that no truncation is evident.  In DDO 154, \citet{carig98}
find no truncation, down to a level of $1 \times
10^{19}$ \cmsq.  This might be indicative of a value for the ionizing background somewhat
lower than that derived for NGC3198, but it is also possible
that the low observed HI
column densities are a consequence of beam dilution; with better
angular resolution, the actual column densities might be observed to
be substantially higher.
In addition, however, as emphasized by Maloney (2001, private communication), the
abruptness of the H~I cutoff depends upon the run of the total hydrogen column density which
is not {\it a priori} known, and the ratio of the neutral to total hydrogen column density
is sensitive to the total hydrogen column density in the regime where the predominate ionization
state switches from neutral to ionized (cf Figures 4 and 8 of Maloney 1993). In fact, the estimate
of the background radiation in NGC3198 was carried out by using the observed H~I column density
to model the total hydrogen column density for assumed values of the ionizing background. It is
the uncertainty of the true radial variation of the total column density of
 hydrogen at these large radii that appears to be the principal uncertainty
in the determination of the background radiation using this technique.

In any event, if the relatively large value of the metagalactic
ionizing flux implied by the HI truncation of NGC 3198 is correct,
then the \Ha\ emission produced by this ionization should be detectable
with achievable sensitivities
in instances where the HI column density is high enough
to ensure that most of the ionizing photons are absorbed.

%%###
For these reasons, we continue to believe that the use of 
 \Ha\ surface brightness
measurements as described in VWRH still provides the route to the
least model-dependent estimate for the metagalactic ionizing flux, {\it provided} that
(i) The targets are relatively isolated from star forming regions and
(ii) that the geometry is well understood, for the reason discussed in
section 3.3
%RJWREVEND 

Consequently, when subsequent further improvements in our
instrumentation were made, we again carried out basically the same
series of observations as described in VWRH, which resulted in
substantial further lowering of an upper limit on the zero redshift
photoionization rate. The purpose of the present paper is to describe
and discuss these observations.

The selected target, \GH, is the same as in VWRH. This large and
isolated cloud of H~I was discovered serendipitously by
\citet{giov89}, and subsequent mapping by \citet{giov91} showed it to
consist of two distinct components---a northeast component (NE) and a
southwest component (SW), with a weaker bridge of H~I connecting the
two. The NE component has embedded in it stars and a small region of
star formation \citep[and references therein]{turner97}. However the
SW component shows no trace of a stellar component \citep[and
references therein]{salzer}. For this reason, we chose to observe only
the SW component since the relatively bright \Ha\ emission from the
hot stars in the NE component would mask the weak diffuse signal we
are attempting to detect. Subsequently, a higher resolution H~I study
(40\arcsec) using the VLA was carried out by \citet{cheng95}; they
showed that the SW component appears to be a flattened inclined disk
with a systematic velocity $V_{hel} = 1260$ \kms\ and a maximum
rotational velocity of $\sim$13 \kms. We discuss the properties of the
SW component further in \S3.3.2. The cocoon of H~I in which the
optical central part of the NE component resides makes it unlikely
that ionizing radiation from the hot stars in this component will
ionize a significant portion of the SW component, and the isolation of
\GH\ from other galaxies makes the SW component an attractive target
for this program.

The plan of the remainder of the paper is as follows: In \S2 we
describe the instrumentation and observational procedures used for the
results described below and in \S3 we describe the reduction
procedures. We present the results in \S4 along with a discussion of
their implications for some of the issues described above which
motivated these observations.

\section{INSTRUMENTATION AND OBSERVATIONS}\label{instrum_observ}

\subsection{Instrumentation}\label{instrumentation}

\subsubsection{Reducing Camera}\label{reducing_camera}

The observations described here were carried out at the Carnegie
Observatories on Las Campanas, using the duPont 2.5m telescope. A
reducing camera was constructed for this purpose. The optical design
was carried out by H. Epps, with mechanical design by A. Schier and
associates.  The design consisted of a collimator with an on-axis beam
size of 70 mm and an effective focal length of 525 mm to match the f/7.5
incoming beam from the Cassegrain focus of the duPont telescope.  A 70
mm Fabry-Perot etalon (whose properties are described in
\S\ref{etalon}) was placed in the collimated beam.  An f/2.5 camera
reimages the telescope focal plane onto a SITe 2048 x 2048 CCD with 24
micron pixels, and a scale at the CCD of approximately 0.78\arcsec\ per
pixel.  The CCD was operated in a mode which yielded a readout noise
of about 6e per pixel. However, as explained below, we were obliged to
accept a tradeoff between adequate sampling of the signal and readout
noise, and thus we binned the image in $4\times 4$ pixels. 
The combined collimator-camera reducing camera is achromatic over a
very broad range and yields an unvignetted field of about 25\arcmin\ in
diameter.

\subsubsection{Etalon}\label{etalon}

The University of Maryland ET-70 etalon was used for this experiment.
The etalon and CS100 controller are made by Queensgate.  The 70 mm
diameter etalon has a gap of 44 $\micron$, and a free spectral range
(FSR) of 48.7 \AA\ at 6590.4 \AA, the expected center of the
redshifted \Ha.  The FWHM of the instrumental response
is 1.40 \AA, and the effective bandwidth (see \S3.1) is 1.64 \AA.  The
etalon was flushed with dry nitrogen during the observations and
calibrations.  The etalon can be tilted so as to direct reflections
out of the field of view.

A four-cavity, 6-inch square, order-isolating filter was located near
the telescope focal plane.  The filter has an intrinsic FWHM of 38
\AA, with a transmission greater than 75\%\ for 6581 \AA\ $< \lambda <
6605$ \AA.  Since the etalon FSR is 48.7 \AA, in this wavelength range
photometric calibration on continuum objects does not suffer from
contamination from other orders.

\subsection{Observations}\label{observations}

We observed \GH\ the nights of 1997 April 7--10 using the reducing
camera and Fabry-Perot etalon described in \S\ref{instrumentation}.
The first three nights were photometric, while the atmospheric
transparency varied during the course of the last night. Consequently,
only data from the first three nights were used in the analysis
described here.

To reach the very low surface brightness required for this
experiment demanded careful attention to observing and reduction
strategy.   Both shot-noise and systematic errors could contribute to
the error budget.   A key strategy for minimizing systematic
errors is to switch as rapidly as possible between the target and sky
positions.  The lower limit on the switching time is set by the need
to minimize the relative error contributed by CCD read noise.
As discussed in \S\ref{instrumentation}, the reducing camera was
designed for this experiment with high throughput, high speed, and a
large field of view.  Nonetheless, due to the narrow effective
bandwidth (1.64 \AA), further steps were required to make the relative
contribution of CCD read-noise to the total error sufficiently small.
First, as noted above we used 4$\times $4 on-chip binning of the CCD; this
yielded 3.1\arcsec\ pixels. Hereafter
we will use ``pixel'' to refer to one of the $4\times 4$ binned pixels. 
Second, all exposures times were 900
seconds in length.

To minimize errors introduced by variations in atmospheric and
instrumental response, we adopted the following observing procedure.
First, we observed blank sky at a position offset in RA to the west
(referred to as W) of the subsequent \GH\ target observation (referred
to as GH1), with the RA offset chosen so that each pair of W and GH1
exposures was observed through very similar azimuth and elevation
ranges.  In this way, we sought to minimize the effects of systematic
errors due to azimuthal and elevation variations in the atmospheric,
telescope, and instrument response.  To eliminate linear temporal
drifts, we repeated an exposure on \GH\
(GH2) followed by an exposure on a sky position offset
to the East (E).  A W-GH1-GH2-E set of four exposures, observed in the
manner described, greatly reduced errors due to directional and
temporal variations in the instrumental and atmospheric response.

We also spatially dithered each exposure by an amount varying from 5 to
45\arcsec\ in order to minimize errors due to stars and other features.

The monochromatic response in the focal plane is along a ring whose
radius $r$ varies according to the usual

\begin{equation}\label{ring_radius}
 \lambda \ = \ \lambda_0 \ (1 + (r/f)^2)^{-1/2} 
\end{equation}

\noindent
where $\lambda_0$ is the wavelength that transmits on-axis and $f$ is
the camera focal length.  As usual, $\lambda_0$ (and hence $\lambda$)
is adjusted by varying the etalon gap spacing, $l$, according to
$n\!\lambda\ = \ 2 \ \mu \ l \ \cos\theta$, where $n$ is the order,
$\mu$ is the refractive index of the nitrogen gas in the etalon gap,
and $\theta$ is the angle with respect to the optical axis.

In Figure \ref{ring_image} we show a typical 900 second exposure
toward the target position.  Due to the large field, non-zero etalon
tilt angle, and the quadratic nature of the wavelength response
(e.g., equation \ref{ring_radius}), three to four orders are present;
i.e.  three to four rings are present for each night sky line.  The
on-axis order is 136, with orders 135, 134, and 133 also present off
the optical axis.  With our large, binned pixels, the quadratic nature
of the wavelength response results in a wavelength shift comparable to
the spectral resolution across a pixel for orders 135, 134, and 133.
Since the wavelength resolution is consequently slightly degraded for
the outer orders, we extract separate spectra for each order.
\notetoeditor{Figure 1 (fig1.ps) is not encapsulated; 
it should go here; the legend is embedded in the text}
\begin{figure}[phtb]
\centering
\epsscale{1.}     % this reduces size of figure
%\plotone{fig1.ps} = ring_image.ps
\caption{
Flat-fielded 900-second exposure toward target position.  The image has
been clipped where the flat-field response drops below 60\%.  Most of
the rings are produced by OH night sky emission; the brighter lines
are labeled.  Note that the same lines appear in each order.  The
brightest line (OH 6577 \AA) is clipped in some places because the
flat-field response near the edge of the filter passband drops below
60\%.  The Galactic \NII\ line also appears. The diameter of the 
field of view is $\sim 25$\arcmin.
}
\label{ring_image}
\end{figure}

By the well known Jacquinot relation, $\Omega R = 2 \pi$ (where
$\Omega$ is the angular area and $R$ is the resolution), the angle on
the sky subtended by a complete monochromatic ring is independent of
radius $r$.  Clearly, we want as much as possible of the ring
corresponding to the redshifted \Ha\ line to be projected within the
\GH\ cloud.  But we also want as much spectral coverage on both sides
of H$\alpha$ as possible.  Coverage at shorter wavelengths extends to
the blocking filter cutoff at 6577 \AA\ and includes the Galactic
[NII] 6584 \AA\ line and the atmospheric OH 6577 \AA\ line.  However,
to observe emission at a given $\lambda$ (in particular around 6590.4
\AA, the wavelength expected for the \Ha\ centroid in \GH\ SW; see
\S\ref{reduction_details}), equation (\ref{ring_radius}) shows that
the maximum wavelength $\lambda_0$ (i.e. the wavelength observed on
axis) increases as $r$ is increased.  In order to keep a substantial
fraction of the 6590.4 \AA\ ring within the cloud boundary, we used a
ring radius of 55--65 pixels (170--200\arcsec). This keeps roughly
half of the ring within the cloud boundary but cuts off the red end of
this spectral order at approximately 6593\AA\ which yields a rather
short range of spectrum to the red of the redshifted \Ha\ over which
to define the continuum.

The etalon and the blocking filter are normally tilted to direct
reflections out of the field of view.  To maximize the ring area on
target, we chose an intermediate tilt angle for the etalon such that
the entire ring at 6590.4 \AA\ lies within the field of view.

As shown by equation (\ref{ring_radius}), the wavelength transmitted by
the etalon varies as a function of position.  The usual method for
constructing a spectrum is to obtain multiple exposures, stepping the
etalon gap $l$ through a full free spectral range, or at least
sufficiently to fully sample the line profile and continuum at all
positions.  However, the long exposure times make this impractical.
Instead, we know that \Ha\ emission ionized by metagalactic radiation
will be nearly uniform in brightness for regions that are optically
thick to the Lyman continuum (as described in more detail in
\S\ref{Halpha_to_jnu}).  Consequently, we can assemble the spectrum
using the varying spectral information provided by the different
pixels in the field of view.  It is essential  that the pixels
used for the \Ha\ portion of the spectrum be limited to those pixels
that lie towards regions of \GH\ that are optically thick to the Lyman
continuum.  Pixels which transmit non-\Ha\ wavelengths are not
restricted to lie towards the cloud.

% Figure 2
\begin{figure}[phtb]
\centering
\epsscale{0.8}     % this reduces size of figure
%\plotone{sky_spectrum.ps}
\plotone{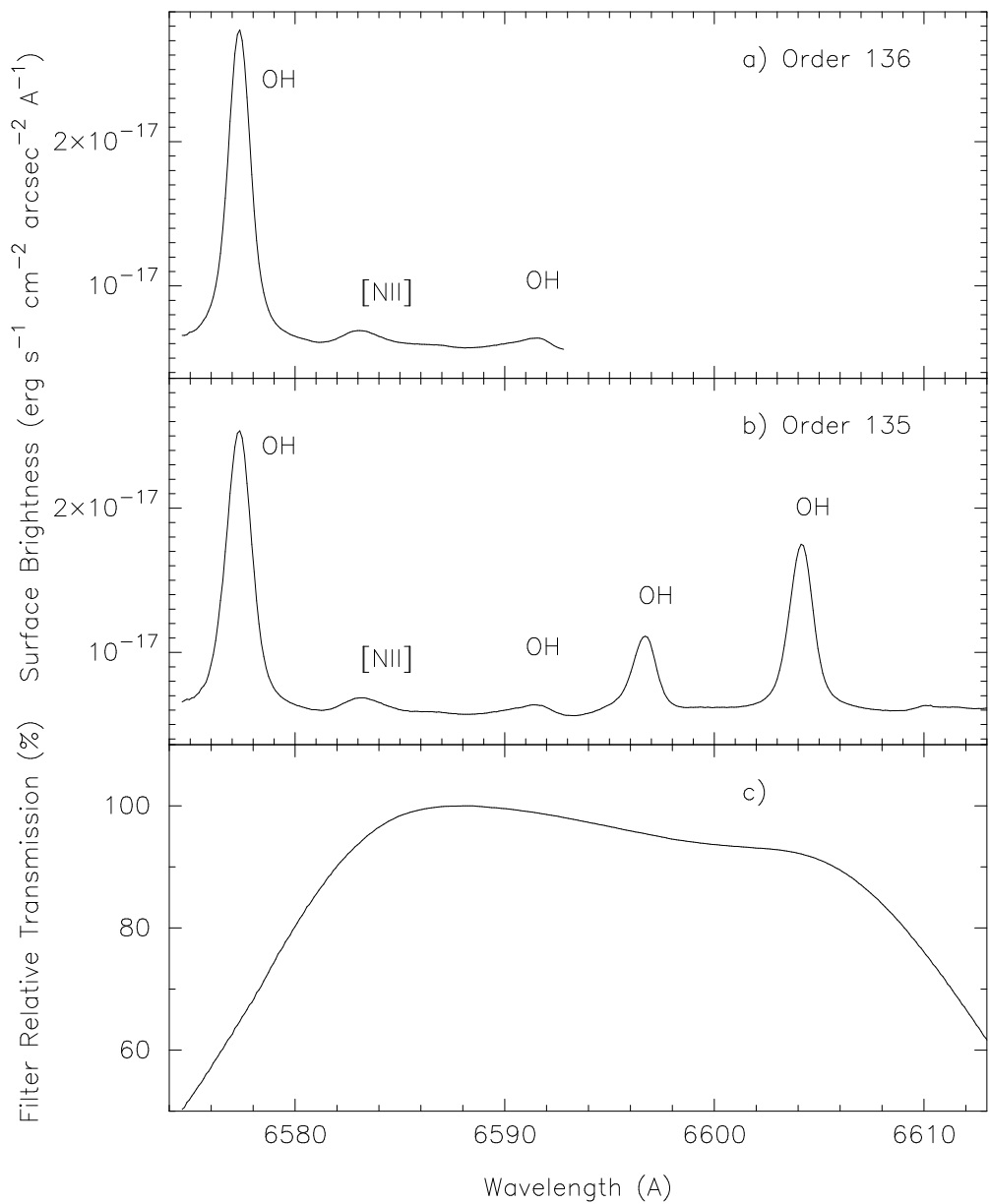}
\caption{
(a) Spectrum for order 136.  Two night-sky OH lines are labeled, along
with the Galactic \NII\ line.  The spectrum is averaged over the
entire field of view for which order 136 is transmitted.  The long
wavelength cutoff occurs because the etalon gap was set so that the
on-axis wavelength was $\sim$6593 \AA.  (b) Spectrum for order 135.
(c) Blocking filter transmission
percentage as a function of wavelength, normalized to the peak
transmission.  The transmission is averaged over the same region as
for the spectrum shown in b).
}
\label{sky_spectrum}
\end{figure}

Figure \ref{sky_spectrum} shows the average sky spectra obtained for
one night for orders 136 and 135, along with the measured transmission
of the blocking filter in the tilted configuration used for the
observations.  Three bright OH atmospheric lines are evident at 6577,
6596, and 6604 \AA, along with a weak OH line near 6592 \AA; only the
6577 and 6592 lines lie within the spectral coverage of order 136.
Also apparent is a bump near 6583 \AA\ corresponding to Galactic \NII.
Continuum due to light scattered from the atmosphere, zodiacal dust,
and Galactic sources is also present at a level of $\sim 6\times
10^{-18}$ \surfperA.

For flux calibration, we observed standard stars each night; the
stars selected included EG 274, LTT 4364, LTT 6248, and LTT 7379,
using equation 3 in \S3.1 to obtain absolute fluxes. 

Most Fabry-Perot programs acquire observations of a diffuse
monochromatic calibration source at intervals of 30 to 60 minutes
during the night to monitor and correct the wavelength response.  Our
program yields high signal/noise ratio spectra of the atmospheric
night sky lines; as described in \S\ref{reduction} these can be used
for wavelength calibration.  Since they are observed simultaneously
with the target, using them provides a more accurate calibration and
also minimizes the need for additional calibration observations during
night time observing.

Daily calibrations were acquired.  These included bias and dark
frames, dome and twilight flats with the etalon gap set to the
spacings used during the night, and neon calibration cubes.  We varied
the tilt angle for the etalon from night to night, and obtained
calibrations appropriate to each night.

\section{REDUCTIONS}\label{reduction}

Due to the narrow effective bandpass, low surface brightness and
relatively small solid angle over which we expect a signal, we are
severely shot-noise limited and in addition must deal with a number of
subtle systematic effects which further degrade the signal.
Therefore, in addition to the observing procedure described above, the
reduction of the data must deal as well as possible with these
difficulties and hence we describe in some detail various steps in the
reduction of the data. We first give a simple overview of the basic
reduction procedure and the approximations implicit in it for both the
flux calibration observations and the target itself.

\subsection{Basic equations }

The count rate in ADU s$^{-1}$ expected from a very small patch of sky
(e.g., a pixel corresponding to solid angle $\Delta\Omega$) may
be expressed as

\begin{equation}\label{response}
(dC/d\Omega)\Delta\Omega \ = \ g\ \epsilon(\lambda_{max}(\Omega)) \  
A_{eff}(\Omega,\lambda_{max}(\Omega)) 
 \int_{-\infty}^{+\infty}
S(\lambda^{\prime},\Omega) \ R(\lambda^{\prime} -
\lambda_{max}(\Omega)) \ d\lambda^{\prime} 
\end{equation}

\noindent
In this expression $S(\lambda,\Omega)$ is the photon surface
brightness per \AA\ as a function of wavelength and direction. At any
given point on the detector, the Fabry-Perot will have its maximum
response at some wavelength $\lambda_{max}(\Omega)$, and at this
wavelength and direction we denote the overall effective collecting
area of the system as $ A_{eff}(\Omega,\lambda_{max}(\Omega))$ (not
including the atmospheric attenuation, $\epsilon(\lambda)$ and the CCD
gain in ADU e$^{-1}$, $g$, which are included separately in equation
(\ref{response})). This effective area includes the spatial and
wavelength dependence of the blocking filter and the pixel-to-pixel
sensitivity of the CCD, determined via flat field images, as described
below in \S3.2.  However, the etalon will have some response in this
direction to photons differing from $\lambda_{max}(\Omega)$ and we
denote this response function relative to its maximum response by
$R(\lambda -\lambda_{max}(\Omega))$. In principle, as in any system
with finite spatial and wavelength resolution, we would need to invert
equation (\ref{response}) to reconstruct the actual distribution of
surface brightness, $S(\lambda,\Omega)$.

In practice, spectra for both the standard star and the target are
constructed as follows.  1) We define a fine grid in wavelength with
a uniform spacing, $\Delta\lambda_{grid}$ of 0.11 \AA. 2) For each
pixel within the target mask (described below) we find the wavelength
$\lambda_{max}$ associated with the center of that pixel, calibrated
using the series of neon lamp observations described below.  3) We
associate all the counts in that pixel with the wavelength bin in this fine
grid closest to $\lambda_{max}$. After extraction on this 0.11 \AA\
interval grid, the spectra were then binned on a coarser 0.34 \AA\ grid, and
all plots, fits and analyses of the spectra use the
0.34 \AA\ grid. We refer hereafter to each
element of this  coarser grid as a ``wavelength bin'' to distinguish it from
the spatial pixels.  

Consider first the observations of an early type standard star for
calibration of the throughput of the system. In this case, over the
wavelength region of interest the flux is constant with wavelength and
the spatial structure is simply the seeing disk. In this case, the
total count rate from the star will be

\begin{equation}
C_{std} \ = \ g\ \epsilon(\lambda_{max}(\Omega)) \
 A_{eff}(\lambda_{max}(\Omega)) \ [dN/d\lambda]_{std} \
 \Delta\lambda_{EB}
\end{equation}

\noindent
In this expression, $ [dN/d\lambda]_{std}$ is the (known)
monochromatic flux (in photons cm$^{-2}$ s$^{-1}$ \AA$^{-1}$), from
which the effective area at this wavelength and this location on the
detector may be calculated. The quantity $\Delta\lambda_{EB}$ is the
effective bandwidth of the etalon, which is the response function
$R(\lambda - \lambda_{max})$ (equation 2) integrated over
$\lambda$. The response function of the Fabry-Perot is quite
accurately represented by a Voigt profile, and we determine the
effective bandwidth empirically using neon comparison lines.  When the
procedure described above is applied to an unresolved continuum
object, the only error is simply that associated with the aliasing of
a detector which is not adequately sampled if the intra-pixel response
is not uniform; however, this effect can be minimized by small
dithering of the standard star observations.
 
For the observation of the target, the opposite situation holds: the
spatial scale is much larger than the pixel size, but the expected
wavelength dependence in the neighborhood of the emission line is, in
our case, comparable to the etalon resolution. In this case, the
count rate from the source if it were finely sampled spatially
can be shown to be just the convolution of the intrinsic emission line
profile with the etalon response:

\begin{equation}\label{extended}
F(\lambda, \ \lambda_0) \ = \ g\ \epsilon \ A_{eff} \  \ B_0 \ 
 \int_{-\infty}^{+\infty}
G(\lambda^{\prime}-\lambda_0) \ R(\lambda^{\prime} -
\lambda) \ d\lambda^{\prime} 
\end{equation}

\noindent
where $B_o$ is the wavelength-integrated photon surface brightness,
and $G$ is the normalized intrinsic line profile.  We have omitted the
angular dependence for clarity.  The instrumental profile $R$ is a
Voigt function, and as discussed in \S\ref{reduction_details}\ the
convolution represented by the integral in equation (\ref{extended})
results in an observed profile which is also well described by a Voigt
function.  Then the photon surface brightness $B_o$ is obtained from
the count rate summed over the line profile $F^\prime = \sum_i F_i \
\Delta\lambda_{grid}$\ as

\begin{equation}
B_o \ = \ { F^{\prime} \over g\ \epsilon \ A_{eff} \
\Delta\lambda_{EB} }
\end{equation}

The effective area $A_{eff} \ = \ \eta \ A$, where $A$ is the
telescope area and $\eta$ is the efficiency of the entire optical path
including telescope and instrument optics, blocking filter, etalon,
and CCD.  Using the standard star observations, we determined that
$\eta = 0.22$.  Photon rates were converted to surface brightness,
corrected for the small amount of Galactic extinction,
which from \citet{schlegel} amounts to about 0.05 mag in the R band.

In practice, as in conventional spectrographs, if the sampling due to
the pixel size is not adequate, there may be aliasing of the
spectrum. For the etalon order nearest the optical axis of the etalon,
the sampling is adequate in spite of the $4\times 4$ binning.  However
for the remaining orders this is no longer the case. If each
wavelength bin were only sampled once, this could result in an
emission line profile whose center and width were poorly defined. In
fact, this situation is entirely mitigated because the format produced
by the etalon causes each bin in the wavelength grid to lie at
more-or-less random positions on the more than 100 pixels contributing
to each wavelength bin, just as ``dithering'' does in a conventional
undersampled spectrograph.

\subsection{Details of the Reduction Procedure}\label{reduction_details}

The first steps in the data reduction for all calibration and data CCD
frames were bias and dark frame subtraction; bias and dark frames were
constructed from the median value for each pixel derived from stacks
of bias and dark frames in order to minimize the noise contribution
from this step.  Next, spatial variations in the instrument and CCD
response were removed by dividing by a normalized flat-field frame
obtained from a stack of exposures of the standard flat-field lamp
illuminating a screen on the side of the dome.  We used a flat-field
frame obtained at the same etalon gap spacing $l$ as used for the
observation; we masked pixels for which the flat-field response was
less than 60\%\ of the peak.

The wavelength response given by equation (\ref{ring_radius})
assumes ideal optics. In practice, there are deviations that result in
slight shifts in the response; for optimum resolution we characterize
these deviations as follows.  We illuminate the dome screen with a
neon lamp, stepping the etalon gap over a free spectral range in
steps of half the wavelength resolution.  Then at each pixel we
determine the etalon gap that maximally transmits the neon calibration
line.  In this way, we characterize the deviations from the ideal
wavelength relation given by equation (\ref{ring_radius}).

As mentioned in \S\ref{observations}, the standard method to monitor
and calibrate temporal variations is to take periodic exposures of a
diffuse monochromatic calibration lamp.  Because we have bright night
sky lines of known wavelength in each exposure, it is more accurate
and efficient to use these.  In principle, we could determine the
centroid and radius of one of the night sky rings; however, in
practice stars and cosmic rays make this difficult.  Instead, we used
the following method.  The wavelength error caused by an incorrect
choice of optical axis and etalon gap varies as a function of position
in the field.  The night sky lines present in the field
(e.g., Fig. \ref{ring_image}) have known wavelengths.  We compare the
known wavelengths with those computed for a large range of assumed
optical axes and etalon gap, and select the axis and gap which yield
the smallest difference between the calculated and known wavelengths.

We used the Tukey biweight statistic as described by \citet{beers90}
to estimate the count rate per pixel for each wavelength bin based on
all the pixels contributing to that wavelength bin. As described by
\citet{beers90}, the biweight estimator is more resistant against
outliers and more robust than the mean, and more efficient than the
median.  It is highly effective at minimizing the effects of cosmic
rays and stars.

To check that variations across the frame do not cause offsets between
the \Ha\ part of the spectrum and the rest of the spectrum, we divided
the frame into several regions and compared the spectra extracted from
each region.  Differences in these spectra were small, and are
dominated by scattered light from stars; this effect is described
next.

Even though spectral features in stars in the field are likely to be
weak, these stars can nonetheless create spectral artifacts.  Recall
that the count rate in each wavelength bin is determined from the
biweight of the count rate for the CCD pixels corresponding to that
wavelength bin.  Although stars are effectively rejected by the
biweight estimator, which excludes pixels with count rates that
deviate by more than four standard deviations from the central
location, scattered light from brighter stars can cause difficulties.
This is because portions of the extended skirts of the point spread
function can be too weak to be rejected by the biweight estimator, yet
extend over a sufficiently large number of pixels so as to bias the
biweight.

Accordingly, we adopted the following procedure.  First, we excluded
regions for which scattered stellar light was detectable by visual
inspection of the frames.  Since the error introduced by a star
increases with the size of the region that escapes the biweight
cutoff, and since the larger regions are of course associated with the
brightest stars, this was straightforward, and in itself greatly
reduced much of the problem associated with star light.  This
eliminated artifacts in individual spectra that were as large as
$5\times 10^{-20}$ \surfperA, although making a difference of less
than $5\times 10^{-21}$ \surfperA\ to the final averaged spectrum.  As
an added precaution, we identified the remaining stars in the field
using the program SExtractor \citep{ber96}.  Next, using a point spread
profile determined on brighter stars, we subtracted the scattered
light contribution from the remaining stars; the resulting changes to
the spectra were minor.

As described in \S\ref{Halpha_to_jnu}, the incident ionizing flux is
inferred from the \Ha\ brightness toward regions with observed column
densities $N(HI) > 10^{19}$\cmsq\ using the 75\arcsec\ H~I maps
provided in digital form by \citet{cheng95}.  Therefore, we excluded
regions with lower H~I column densities in constructing the spectrum
for the wavelength interval expected for the redshifted \Ha\ emission,
6588.9 \AA\ $< \lambda <$ 6591.9 \AA. Because pixels with
wavelengths outside this range were not excluded, the number of pixels
per wavelength bin over the wavelength interval defining the continuum
was a factor of two to three higher than for the region of line
emission.

In Figure \ref{wcce_image} we show the image resulting from directly
adding the GH1 and GH2 frames of one W-GH1-GH2-E set and subtracting
the W and E frames.  The OH night sky lines largely cancel, while the
\NII\ line appears black, corresponding to brighter \NII\ emission in
the GH1 and GH2 frames than the sky frames.  Note the patchy spatial
distribution of the \NII\ emission and the presence of two well
defined velocity components.  Also note that the OH line residuals
appear to have a slight offset in the subtracted image.  This occurs
because of drifts in the etalon.  Because of such drifts, spectra are
extracted not from W-GH1-GH2-E combined frames such as shown here, but
from the individual frames because drifts in the wavelength scale can
be corrected during the extraction process, as described above.
The ring corresponding to the approximate location of the expected
redshifted \Ha\ signal with centroid  at 6590.4 \AA\ is shown by the solid-line circle.
\notetoeditor{Figure 3 (fig3.ps) is not encapsulated; 
it should go here; the legend is embedded in the text}

\begin{figure}[phtb]
\centering
\epsscale{1.}     % this reduces size of figure
%\plotone{fig3.ps} = wcce.ps
\caption{ Target minus sky frame produced from a single W-GH1-GH2-E
set.  Several OH and \NII\ features are marked.  Spatial variations in
the Galactic \NII\ features are evident, corresponding to differences
in \NII\ emission velocity and spatial distribution between the target
and sky fields, along with OH night sky residuals.  Also indicated are
the $N(H) = 10^{19}$ \cmsq\ contour for the southwest \GH\ cloud (GH
SW) and for the bridge between the SW and NE components; the H~I data
are from \citet{cheng95}. The cross marks the etalon optical axis. The
solid circle shows the locus of 6590.4 \AA\ in order 136, the expected
wavelength for \Ha, while the dashed arc shows the same locus for
order 135, where little or no \Ha\ is expected.  }
\label{wcce_image}
\end{figure}

\begin{figure}[phtb]
\centering
\epsscale{0.8}     % this reduces size of figure
%\plotone{all_spectra_136.ps}
\plotone{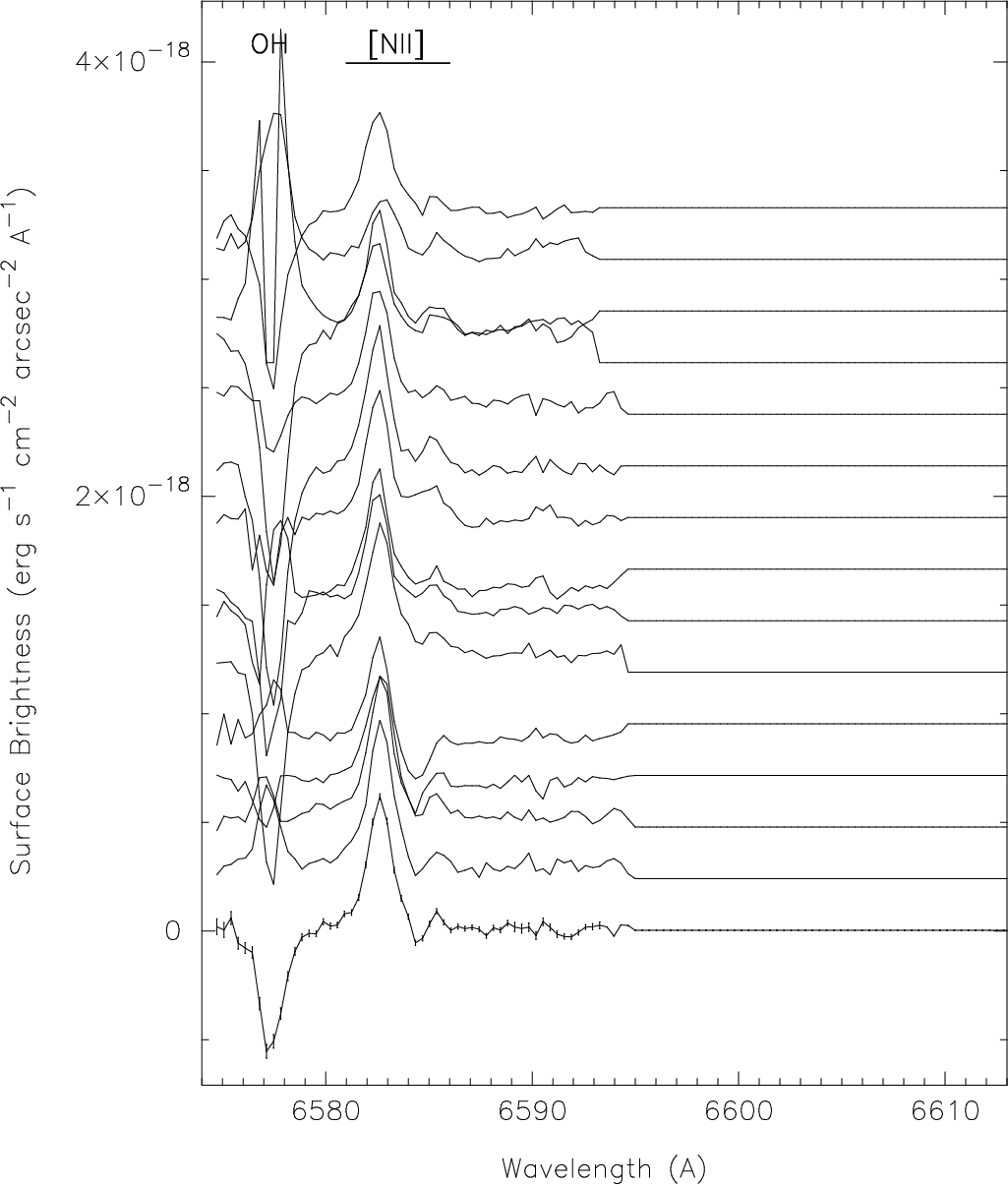}
\caption{ 
Target minus sky spectra for each of the 15 W-GH1-GH2-E sets
obtained during the run are shown for order 136. The error bars shown
on the bottom spectrum are calculated from Poisson statistics.  
}
\label{all_spectra_136}
\end{figure}

\begin{figure}[phtb]
\centering
\epsscale{0.8}     % this reduces size of figure
%\plotone{all_spectra_135.ps}
\plotone{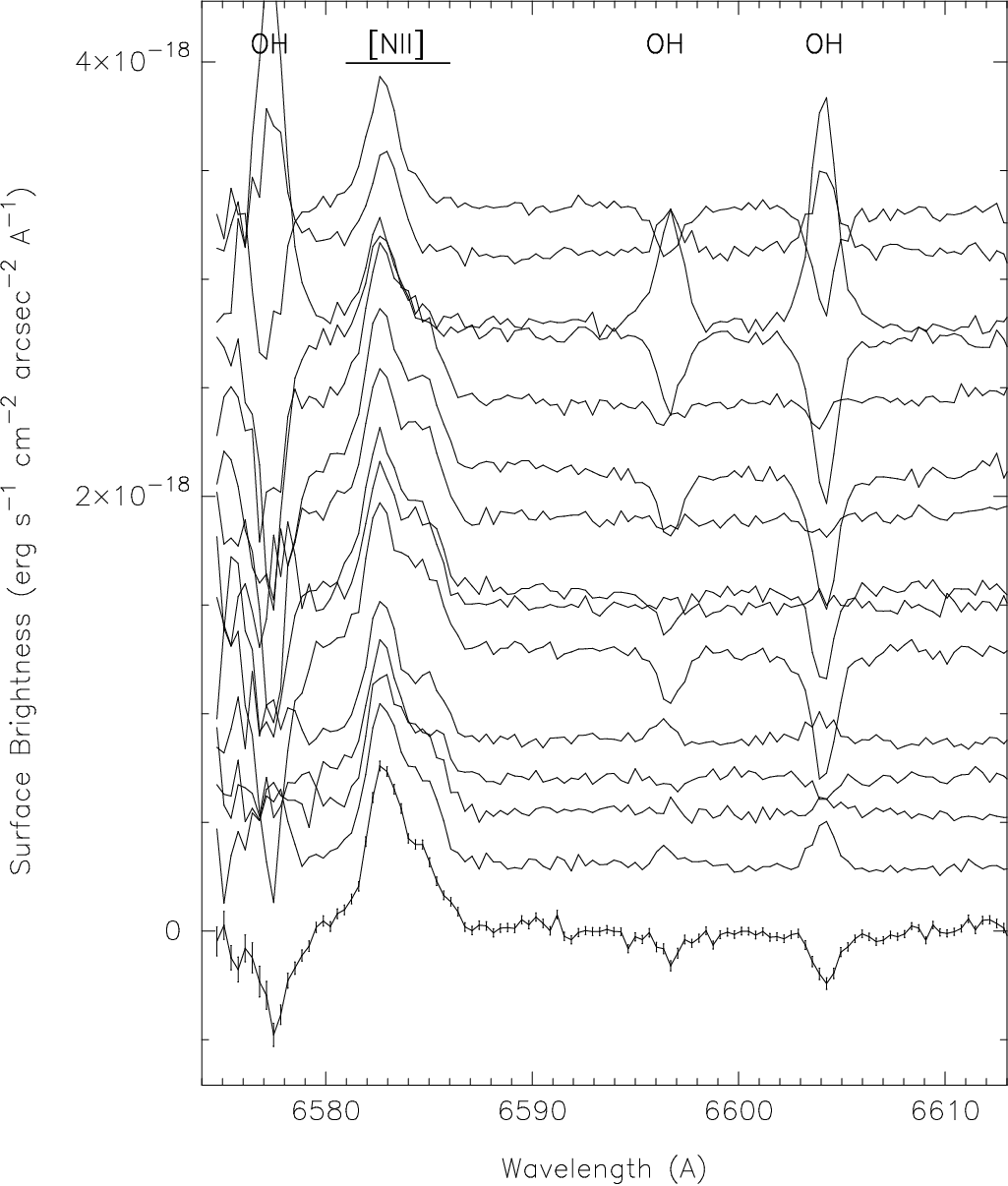}
\caption{
Target minus sky spectra for each of the 15 W-GH1-GH2-E sets obtained
during the run are shown for order 135. The error bars shown on the bottom
spectrum are calculated from Poisson statistics.
}
\label{all_spectra_135}
\end{figure}

Target minus sky (i.e. On-Off) spectra produced for orders 136 and 135
from each W-GH1-GH2-E set of four spectra are shown in Figures
\ref{all_spectra_136} and \ref{all_spectra_135}; each On-Off spectrum
corresponds to 30 minutes integration time on the target.  The error
bars shown on the bottom spectrum of Figures \ref{all_spectra_136} and
\ref{all_spectra_135} are calculated from Poisson statistics; the
errors are larger at the wavelengths of night sky lines due to the
larger count rates, and in the wavelength range 6588.9 \AA\ $< \lambda <$
6591.9 \AA\ due to the smaller number of unmasked pixels. Significant
non-zero residuals are seen at the wavelengths of the Galactic \NII\
line; this is due to the difference in high Galactic latitude \NII\
emission between the on and off positions.  Also the atmospheric lines
and the continuum level typically show non-zero residuals due to
temporal and spatial variations in night sky brightness.

\begin{figure}[phtb]
\centering
\epsscale{0.8}     % this reduces size of figure
%\plotone{all_averages.ps}
\plotone{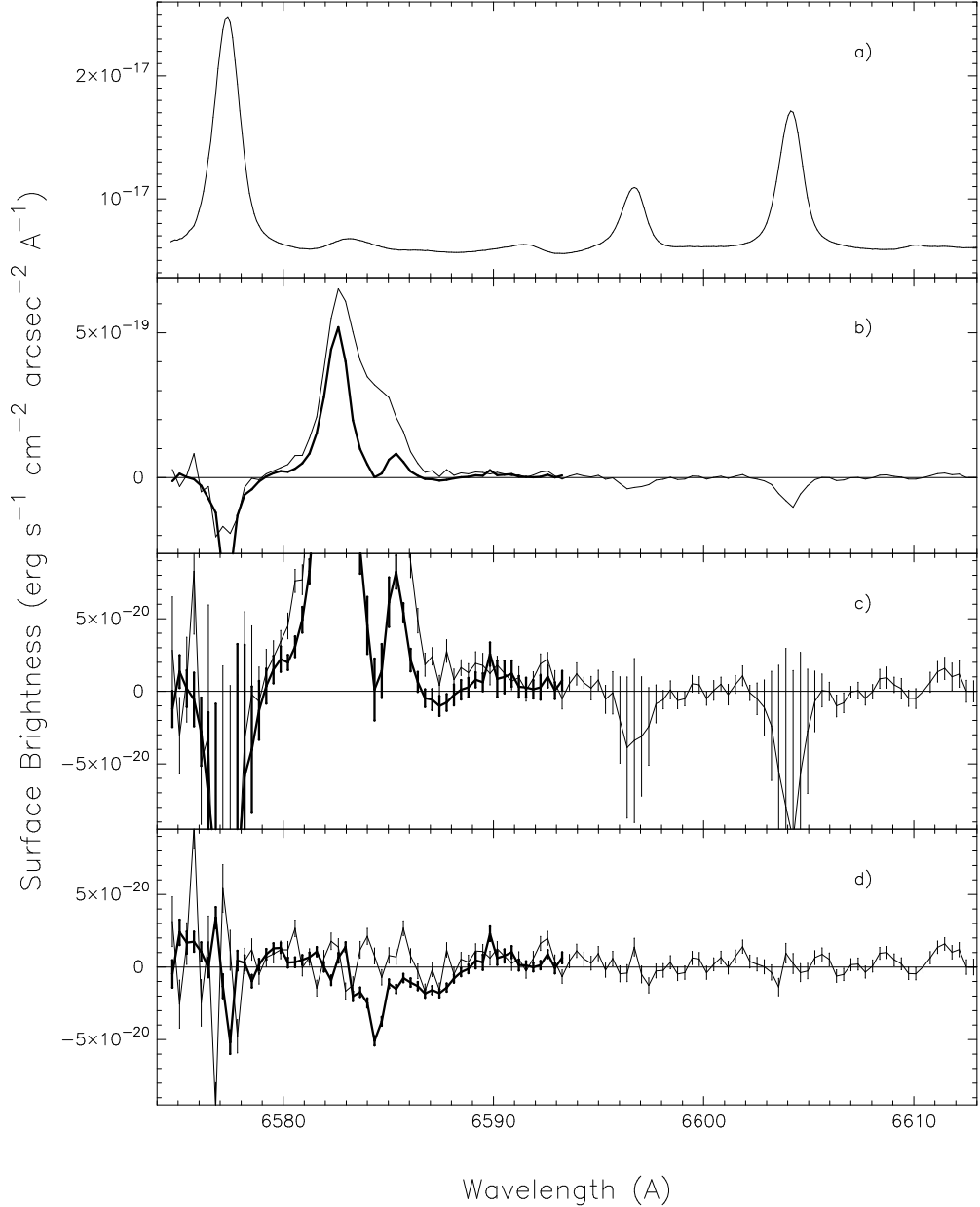}
\caption{ (a) Same spectrum as in
Figure \ref{sky_spectrum}b showing the lines in the bandpass.  
(b) The average of all target minus sky
spectra for order 136 (thick line) and 135 (thin line).  Multiple
components in the residual \NII\ line are evident, due to differences in the
spatial distribution of \NII\ between the target and sky positions.
(c) Expanded view of spectra shown in the panel above.  Error bars are
1$\sigma$\ derived from the scatter of the 15 W-GH1-GH2-E spectra.
(d) The same spectra as
in (c), except the continuum, OH, and \NII\ residuals have been
subtracted as described in the text.  Here the error bars are those
calculated from Poisson statistics.  }
\label{all_averages}
\end{figure}

Figure \ref{all_averages}a is reproduced from Figure
\ref{sky_spectrum}b to enable comparison with Figure
\ref{all_averages}b which shows the target minus sky spectra for
orders 135 and 136 averaged over the 15 W-GH1-GH2-E spectra obtained
during the run.  In Figure \ref{all_averages}c, we show a magnified
view of the spectra, including error bars.
The error bars in Figure \ref{all_averages}c are estimated from the
scatter at each wavelength between the 15 spectra that contribute to
the average and have decreased by approximately a factor of four from
the individual W-GH1-GH2-E spectra, as expected.  Small negative
residuals remain at the OH wavelengths, at a level of about 1\%\ of
the line strength, due to nonlinear temporal variations in the OH line
strength.

There is also a non-zero residual continuum level.  We determined this
residual continuum level (assumed constant with wavelength over this
narrow range) by averaging over 6598.4 \AA\ $< \lambda <$ 6601.5 \AA\ and
6606.3 \AA\ $< \lambda <$ 6610.8 \AA\ for order 135.  The residual
continuum level is $1.4\times 10^{-20}$ \surfperA, about 0.3\%\ of
the continuum brightness.  As expected the percentage continuum
residual is smaller than the OH residual since the OH lines are
intrinsically more variable.

There is a weak OH line at 6591.7 \AA\ evident in Figure
\ref{all_averages}a and which also has been identified by Osterbrock
et al. from compilation of Keck spectra \citep{deo97}.  We fit each of
the four OH lines in Figure \ref{all_averages}a to determine the
ratios of the OH line intensities.  The 6591.7 \AA\ line is about
10\%\ of the strength of the 6596 \AA\ line, similar to what can be
derived from the \citet{deo97} Keck data.  Based on the residual negative
OH emission of the bright OH line at 6596 \AA\ of about $2 \times
10^{-20}$ \surfperA, we expect a peak residual of approximately $-2
\times 10^{-21}$ \surfperA\ for the 6591.7 line.  We fit the residual
fluxes in each of the OH lines shown in Figures \ref{all_averages}b
and \ref{all_averages}c using a Voigt function; for the fit, the
Gaussian and Lorentzian line width parameters were held fixed using
values determined from a fit to the neon calibration line.  Also, the
wavelength was fixed at the known value of the OH lines, and the
continuum was fixed at zero since the fits were made to continuum
subtracted data.

The residual \NII\ peaks in Figure \ref{all_averages}b are at $-$50
\kms\ and $+$80 \kms\ respectively. The brightness of this \NII\
feature is close to the detection threshold of the observations of
known high velocity clouds made with the same instrument \citep{ben},
but with the sensitivity of the present observations it is a highly
significant detection and clearly not an artifact.  The brightness of
the 80 \kms\ feature varies significantly on scales of roughly an arc
minute, and arcs corresponding to this feature can even be seen in a
few locations in Figure \ref{ring_image}.

In attempting to subtract off the \NII\ from the spectrum it is
necessary to model the feature with both components due to the broad
Voigt wings in the etalon point spread function. Unfortunately,
however, the strength and extent in velocity space of this \NII\ makes
the locating of the continuum around the expected position of the
redshifted \Ha\ more difficult, since the \NII\ emission is clearly
present to within $\sim140$ \kms\ of the \Ha\ and conceivably could
extend to greater velocities at lower intensity. Moreover, the clear
differences in structure of the \NII\ between the on and off positions
makes the on-off sky subtraction problematic over the spectral region
around the \NII. Note that the \NII\ spectra even over the relatively
small angles separating orders 136 and 135 are slightly different due
to spatial variations in the \NII.  These can readily be seen by direct
inspection of the difference images (e.g., Figure \ref{wcce_image}).
 
Although the presence of the \NII\ adversely affects the measurement
of the \Ha\ signal which is our primary objective, it nonetheless is
an interesting feature in itself.  We are not aware of other
observations in this general part of the sky that would have the
sensitivity to detect \Ha\ or \NII\ emission at the level we detect.
In H~I, the Leiden-Dwingeloo survey \citep{bh94} clearly detects
emission only in the range $-50 < V_{LSR} < +40$ \kms. A similar
result is found in the HIPASS survey \citep{barnes}. Thus, we cannot
at present associate the 80 \kms\ \NII\ emission with known H~I
emission.  The velocity of the 80 \kms\ feature is nearly large enough
to qualify as a high velocity cloud \citep{wvw}; however, \citet{wvw}\
emphasize that the deviation velocity (the difference from the largest
velocity permitted by Galactic rotation) is a better criterion,
arguing for a deviation velocity greater than 50 \kms\ as the cutoff.
Using the model presented by \citet{wakker}, positive velocities due
to rotation are not expected in this direction due to its high
Galactic latitude ($b$ = 64\arcdeg); consequently the deviation
velocity is 80 \kms, suggesting that the detected emission may be
produced by a high-velocity cloud.  With an emission measure of 25 mR,
the \NII\ emission is comparable to the weakest \Ha\ emission from
known high velocity clouds observed by \citet{ben}; however, \NII\
emission was not detected by \citet{ben}\ toward the faintest high
velocity clouds, with upper limits somewhat below our detection.  In
other words the \NII\ emission we detect is somewhat brighter than in
some known H~I high velocity clouds, in spite of the fact that
no H~I emission is seen in this direction at this velocity.
The detected \NII\ emission is
clearly interesting and merits further study to determine its origin.

The resulting spectra with the continuum, OH line, and \NII\ line
residuals subtracted are shown in Figure \ref{all_averages}d. In this
figure, the error bars have been calculated from Poisson statistics. Note
that the order 135 and 136 spectra are generally the same within the
statistical errors, except for part of the \NII\ line; this is due to
spatial structure in the \NII\ distribution and the fact that the
difference in \NII\ distribution between the target and sky positions
must be slightly more complex than can be modeled with just two
components.  

Next we fit a Voigt profile to this spectrum over the wavelength
interval 6588 \AA\ $< \lambda <$ 6593.5 \AA\ (corresponding to about four
spectral resolution elements).  The continuum is fixed to be zero
since the continuum has been subtracted, as discussed above.  The
instrumental profile is taken from the neon calibration.  

\begin{figure}[phtb]
\centering
\epsscale{0.8}     % this reduces size of figure
%\plotone{HI_prof.ps}
\plotone{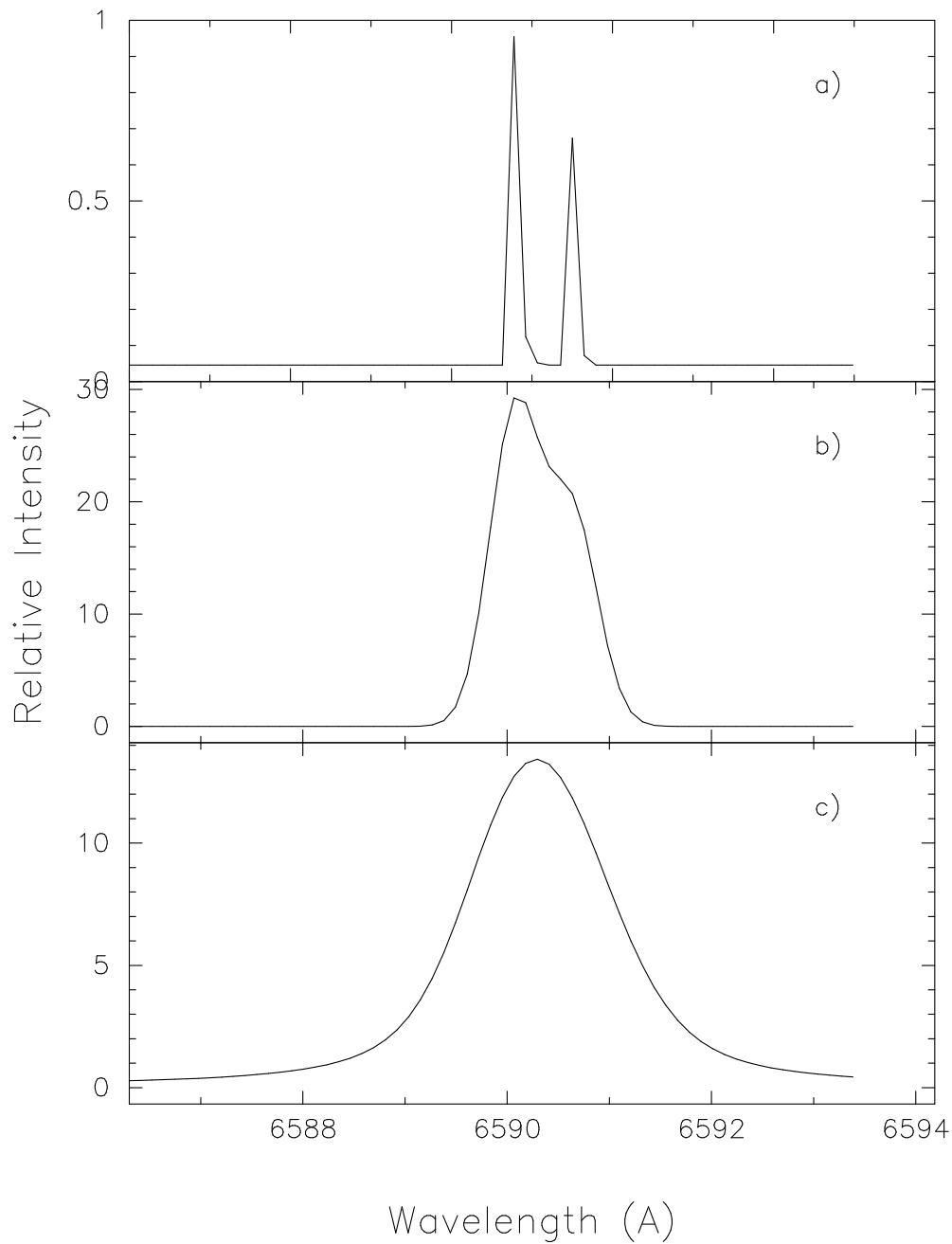}
\caption{
(a) Wavelength distribution of redshifted \Ha\ centroids expected
based on 75\arcsec\ VLA maps of H~I obtained by \citet{cheng95} for
\Ha\ emission regions observed with the Fabry-Perot.
(b) Expected intrinsic \Ha\ profile obtained by convolving the
H~I profile in (a) with Gaussians for H atoms at 10000 K and the H~I
velocity dispersion.  (c) Expected observed \Ha\ profile, obtained by
convolving the intrinsic \Ha\ profile in (b) with the Fabry-Perot
instrumental profile.
}
\label{HI_prof}
\end{figure}

The assumed
intrinsic \Ha\ line profile was obtained as follows. 
Using the  75\arcsec\ resolution VLA H~I observations of \citet{cheng95} 
(which are given in the heliocentric frame), we determined the H~I velocities
of the gas at each pixel on the Fabry-Perot ring at the approximate
redshifted wavelength of \Ha\ and then shifted these velocities by +5.8
\kms, which puts them in the barycentric frame which we have used
throughout this work.
Weighting each pixel equally
as a first order approximation since the \Ha\ emission is only weakly
dependent on H~I column density for the observed column density range,
we constructed the spectrum shown in Figure \ref{HI_prof}a. H~I
emission occurs mostly at two velocities because \GH\ has a rotation
curve characteristic of a galactic disk, and the 6590.4 \AA\ ring has
relatively little overlap with the central part or minor axis
(e.g., Fig. \ref{wcce_image}).  
This profile was then convolved with a Gaussian with a dispersion of
6 \kms\ (corresponding to a typical dwarf dispersion and slightly less
than the velocity dispersion of 8 \kms\ of the NE component) and also
with a Gaussian whose width corresponds to the thermal velocity of
the ionized hydrogen at a temperture of 10000 K. This produces the
profile shown
in Figure
\ref{HI_prof}b. Finally, this profile was convolved with the
instrumental profile to produce the profile
shown in Figure \ref{HI_prof}c.  This corresponds to the profile of
the \Ha\ line we expect to observe from \GH.  
This is well fit by a Voigt profile centered at 6590.4 \AA\ and FWHM 
79 \kms.

\begin{figure}[phtb]
\centering
\epsscale{0.75}     % this reduces size of figure
%\plotone{voigt_fit.ps}
\plotone{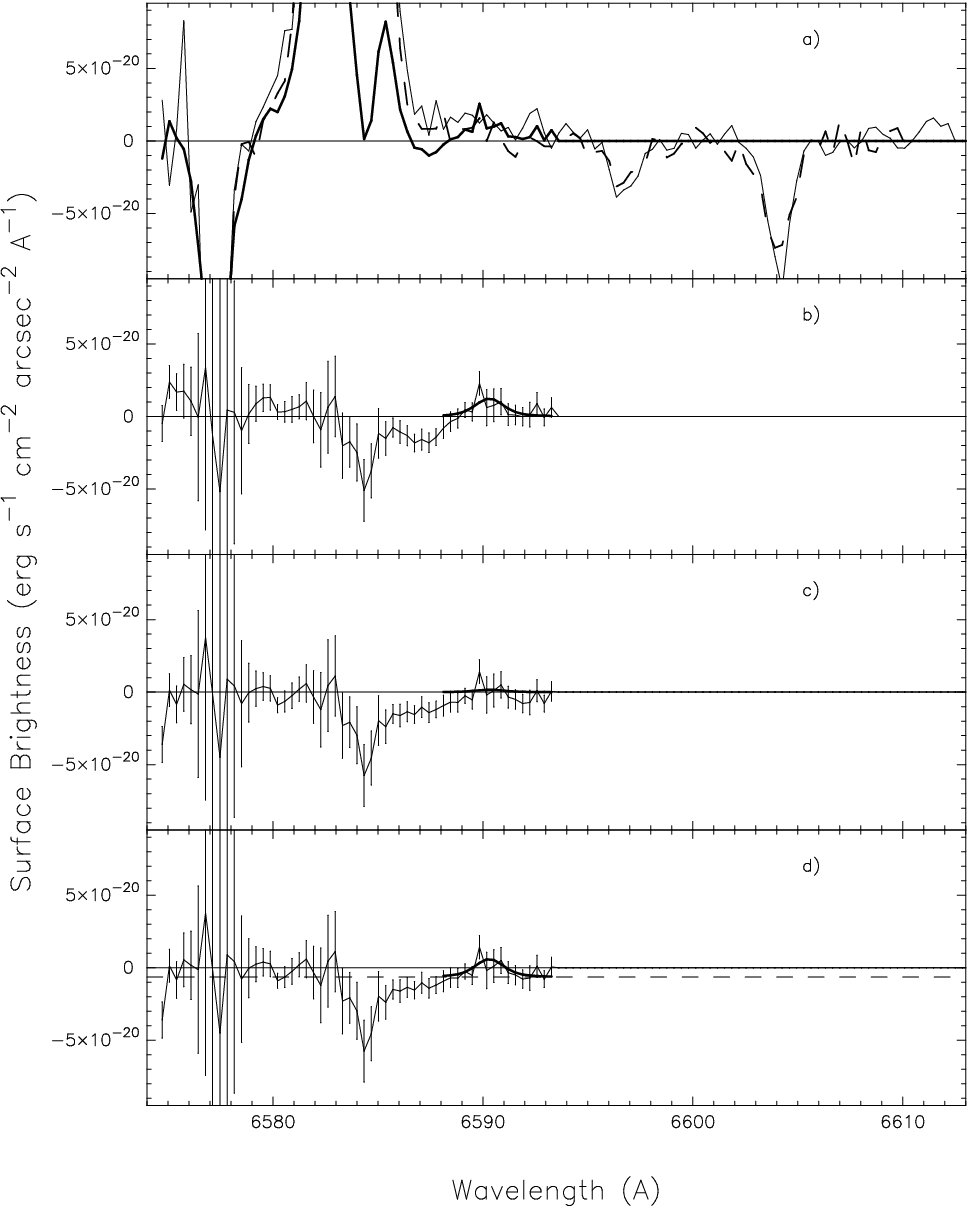}
\caption{ (a) Target minus sky spectra compared for orders 136 (thick
solid line), 135 (thin solid line), and 134 (dashed line).  In this
panel, unlike those below, OH and \NII\ residuals have not been
subtracted. 
 (b) Order 136 spectrum with continuum, OH and \NII\
lines subtracted, shown as a thin line; the error bars are estimated
from the data.  The thick line is a fit of a Voigt profile
in which the only free parameter is the line strength.  The
best fit emission line has strength 4.2 $\pm$ 1.6 mR.  (c) As in (b),
except a smoothed order 135 spectrum has been subtracted. 
The Voigt fit  has strength 0.8
$\pm$ 1.6 mR.  (d) As in (c) except that the fit has both line strength and 
continuum level as free parameters.  The best fit is 3.9 mR $\pm$ 2.2 mR.  The best-fit
continuum is shown as a horizontal dashed line.  
}
\label{voigt_fit}
\end{figure}

A Voigt profile with this shape is then used as the line profile in a
fit to the observed spectrum, in which the only free parameter is the
Voigt line strength.  The best fit line brightness is 4.2 mR.  The
spectrum and the 4.2 mR best fit for the OH, [NII], and continuum
subtracted spectrum for order 136 are shown in Figure
\ref{voigt_fit}b.  To estimate the uncertainty in the line strength,
we used a Monte Carlo approach \citep{benmonte}, in which we
constructed spectra using the errors estimated from the comparison of
the individual W-GH1-GH2-E spectra (i.e. the errors shown in Figure
\ref{all_averages}c).  The resulting 1$\sigma$\ error bars are $\pm
1.6$~mR, and the 2$\sigma$ upper limit is 7.4 mR.

Although we have attempted to accurately apply flatfield corrections
to the data, differences in the spectral and spatial properties
between the flatfield lamp and the sky data, as well as differences
between the resolution in orders 135 and 136 mean that direct
subtraction of the ``on--only'' data in order 135 from 136 to obtain
the net signal in the region of the expected \Ha\ signature may well
mask the signal we are attempting to detect. For this reason we have
employed the on-off chopping described previously, which should
accurately remove the effects of these differences in the instrumental
response.

However, although the sequence of on-off observations removes these
differences as well as temporal variations in the night sky background
which are linear, the characteristic time for the full cycle for one
off-on-on-off sequence is over an hour, and variations over a shorter
time may well occur. Therefore the order 135 spectrum of Figure
\ref{all_averages}d was subtracted from order 136 in the same figure
to produce Figure \ref{voigt_fit}c.  (In order to reduce the noise in
this subtraction the data in order 135 were first smoothed using a
5-point running mean.)

Disappointingly, the signal present in Figure \ref{voigt_fit}b appears
to have largely disappeared.  
It is 
known that other weak night sky emission features are present.  In
particular, weak emission from O$_2$ has been found in the 
night sky spectrum  \citep{slangero2}
and one of these emission features falls at the expected
wavelength of \Ha\ (T. Slanger 2001, private communication).
No attempt has been made to
model any residual emission from this source.
However, the impression is that the main
effect of the subtraction of 135 from 136 is simply to lower the level
of the continuum so that it falls below the nominal zero level. This
impression is confirmed by again fitting a Voigt profile to the same
data as shown in Figure \ref{voigt_fit}c, but now allowing the
continuum level to be a second free parameter. The resulting best fit,
shown in Figure \ref{voigt_fit}d, yields 3.9 mR, very similar to that
found above. Because of the very limited continuum region in order
136, which is affected nearly up to the \Ha\ signal at shorter
wavelengths by the [NII], as described above, and only extends to
$\sim~6593$ \AA\ which is the termination of order 136 as explained in
\S2.2, it is not clear whether such an offset really exists.
  
\begin{figure}[phtb]
\centering
\epsscale{0.8}     % this reduces size of figure
%\plotone{two_sigma.ps}
\plotone{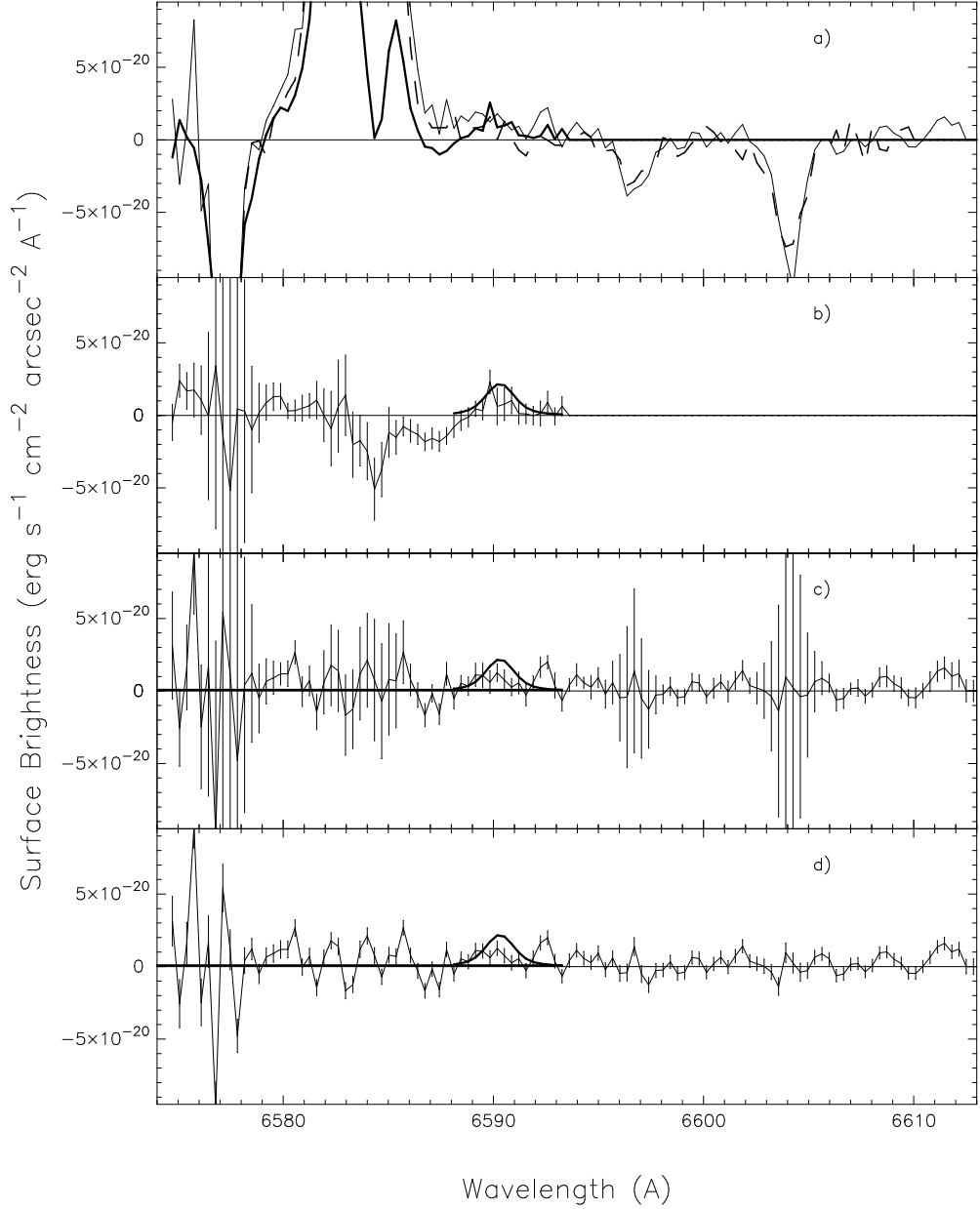}
\caption{ (a) Target minus sky spectra compared for orders 136 (thick
solid line), 135 (thin solid line), and 134 (dashed line).  In this
panel, unlike those below, OH and \NII\ residuals have not been
subtracted.  (b) Order 136 spectrum with continuum, OH and \NII\
residuals subtracted, shown as a thin line; the error bars are
estimated from the data.  The thick line shows a 7.4 mR emission line,
corresponding to the 2$\sigma$ upper limit.  (c) Order 135 spectrum,
with 7.4 mR emission line overlaid.  (d) Same as (c), except the error bars are
Poisson.}

\label{two_sigma}
\end{figure}

It is also not clear what could give rise to such a continuum offset. Spatial
gradients in the background continuum arising from the zodiacal light
and airglow continuum\footnote{This airglow is generally ascribed to
NO$_2$, although there is some disagreement on this point.}  would
have to be present over an extent of only $\sim$8\arcmin.    It should be borne in
mind however that the possible offset is only about 0.1\%\ of the
total continuum level.  The data provide no evidence for such
gradients; first, the CCD dewar was rotated 180\arcdeg\ after the
second night, yet comparison of spectra obtained from the opposite
rotations does not show mirrored spectra artifacts; second, comparison
of the order 135 and 134 spectra in Figure \ref{two_sigma}d reveals no
significant difference, even though these are extracted from regions
offset by $\sim 5$\arcmin, nearly as large as the 7--10\arcmin\ offset
between orders 136 and 135.

A more likely possible source of artifacts is scattered light from
stars that has not entirely been removed or masked.  This scattered
light would be more likely to create spectral artifacts in the central
order.  Partly this is because out of focus reflections between the
etalon and blocking filter are not directed out of the beam for the
central order due to the etalon tilt angle selected.  Also, due to the
quadratic nature of the wavelength dispersion scattered star light
affects relatively narrow spectral regions in the central order
compared to the outer orders.  This may explain why the order 135 and
134 spectra (e.g., Figs. \ref{voigt_fit}a) are quite similar, and more
featureless than the order 136 spectrum.  

Although there is some weak indication of the detection of a line from
Figures \ref{voigt_fit}b and \ref{voigt_fit}d, with strengths of
$\sim$ 4 mR, because of this uncertainty we cannot claim a positive
direct detection of the ionizing background radiation, but the formal
2$\sigma$ upper limit\footnote{The Monte Carlo calculations actually
lead to a direct estimate of an upper confidence limit of 97.5\%,
which for a Gaussian distribution would imply that 95\% of the Monte
Carlo trials lie between -2$\sigma$ and +2$\sigma$ of the best fit
value, but we will simply refer to this as the ``2$\sigma$'' upper
limit.} to any signal from the fits in either 8b or 8d is about 8 mR,
and we shall use this limit in our subsequent discussion. The actual
formal upper confidence limits to the fits in Figures
\ref{voigt_fit}b, c, and d are shown in Figures \ref{two_sigma}b,
\ref{two_sigma}c, and \ref{two_sigma}d, where Figure \ref{voigt_fit}a
is again reproduced for reference as Figure \ref{two_sigma}a.

\par
\noindent
\subsection{Conversion of the \Ha\ Surface Brightness to Ionizing
Flux}\label{Halpha_to_jnu}

Following VWRH and \cite{sto91}, (see also SRGPF) the one-sided
incident ionizing flux is defined as

\begin{equation}
\Phi = \pi \! \int_{\nu_0}^\infty {J_{\nu}\over h\nu}\, \, d\nu 
\end{equation}

\noindent
where $J_\nu$ is the mean intensity. The factor $\pi$ assumes that the
background intensity is isotropic. From our measurement of the \Ha\
surface brightness, $I_{H\alpha}$, we may calculate $\Phi$ from:

\begin{equation}\label{phi_ex}
\Phi  =
4 \pi \, \, {I_{H\alpha} \over h \nu} \, \, {1 \over f_a \, 
           f_{H\alpha} }
              \, \, \biggl( {A_{proj}\over A_{tot}} \biggr)
\end{equation}

\noindent
which can be written as

\begin{equation}\label{phi_ex_calc}
\Phi = (1.0\times 10^3 {\rm ~~cm}^{-2} {\rm s}^{-1}) \, \,
  \biggl( {I_{H\alpha} \over mR} \biggr) \, \, {1 \over f_a  \,
  f_{H\alpha} } \  \biggl( {A_{proj}\over A_{tot}} \biggr)
\end{equation}

\noindent
In equations (\ref{phi_ex}) and (\ref{phi_ex_calc}), $f_a$ is the
fraction of the one-sided incident flux which results in
recombinations to excited states, $f_{H\alpha}$ is the fraction of
those recombinations which produce \Ha\ photons, $A_{tot}$ is the
total surface area exposed to the incident radiation and $A_{proj}$ is
the observed projected surface area. This expression is quite general,
except that for complex geometries the appropriate averages over the
cloud surface of the quantities $f_a$ and $f_{H\alpha}$ need to be
taken, and the $I_{H\alpha}$ surface brightness is that averaged over
the projected area.

\subsubsection{Effective $H\alpha$ Recombination Rate}
The \Ha\ spectrum has been extracted from regions for which $N_{HI}
\geq \, 10^{19} \, {\rm cm}^{-2}$. However, as described in section 3.3.2,
recent work by \citet{cheng95} suggests that \GH\ SW is a disk which is
viewed far from the normal, with an aspect ratio of 4:1, so that the
ratio of the total area to the projected area is $\sim8$. If for
simplicity we assume plain slab geometry, then the column density
{\it normal} to the disk corresponding to the $10^{19} \, {\rm cm}^{-2}$
column density is $2.5 \times \ 10^{18} \, {\rm cm}^{-2}$, corresponding
to an optical depth at the Lyman limit of 15.7. Adopting the spectral
index of 1.8 suggested by SRGPF we find that at this optical depth about 87\% of the
normally incident photon flux is absorbed within the disk. However,
some of the resulting recombinations occur to the ground state, and a
portion of this diffuse Lyman continuum radiation can escape from the
slab after one (or more) such scatterings, and thus does not
contribute to any Balmer radiation.
The fraction of the recombinations which go to the ground state
depends upon the electron temperature, as does the energy distribution
of this recombination radiation, and as discussed below, we adopt an
electron temperature of 10000 K. 
Since most of the region around the Fabry-Perot 6590.4 \AA\
ring consists of H~I column densities in excess of $10^{19} \, {\rm cm}^{-2}$,
we calculate the average value of $f_a$ by creating a histogram of column
densities which contribute to the ring (after reducing the observed values
by the geometric factor of four) and for each of the values of the 
column density in the histogram compute the value of $f_a$, weighting
each of these values by the number of pixels contributing to that bin.
Detailed calculations then yield a
fraction of the normally incident photon flux which ultimately results
in recombinations to excited levels of 0.88, which is the quantity
$f_a$ in equations (7) and (8). (Had we used a spectral index
of 1.0, the number would be reduced to 0.75. We also note that 
even for a normal column density 10
times lower than our minimum value the fraction of the continuum flux which results in
excited state recombinations is still slightly greater than half
of the 0.88 value, namely 0.42.)
 
We also require the effective recombination rate for producing \Ha\
photons, or in the formulation used in equations (7) and (8), the
fraction $f_{H\alpha}$ of excited state recombinations which ultimately
produce an \Ha\ photon. The value of $f_{H\alpha}$ depends, as does
$f_a$, upon the electron temperature, and upon whether Lyman line
photons readily escape following the recombinations (``case A'') or
scatter internally until they are ultimately converted to Balmer and
higher series photons (``case B''). Since nearly all of the
recombinations occur at optical depths in the Lyman lines which are
very large, case B is appropriate.

In VWRH we adopted a temperature of 20000 K on the assumption that a
hard spectrum and low metallicity would lead to an electron temperature
much higher than the value of $\sim10000$ K usually assumed.  However
this assumes a balance between atomic heating and cooling
mechanisms. In regions of the low density intergalactic medium
producing the low column density \Lyf\ lines, \citet{dave} deduce much
lower temperatures ($\sim5000$ K) due to the fact that Hubble expansion
dominates over atomic cooling processes. However, these considerations
may not be applicable to the material comprising \GH\ where the
expansion has evidently been halted, and where the particle density is
probably much higher than that in which the low column density
Ly$\alpha$ lines arise, in which case thermal equilibrium is a much
better approximation.  
Thermal equilibrium calculations for a range of metallicities and
particle densities for gas exposed to metagalactic radiation and
partially shielded by neutral H~I yield temperatures in the range 
8000--10000 K (Wolfire 2001 private communication). However his calculations do not extend below column
densities of $10^{18} \, {\rm cm}^{-2}$ so that on the outer skin of the
cloud the temperatures might be slightly higher.  We adopt an electron
temperature of 10000 K.  Then, for case B and an electron temperature
of 10000 K from data in \citet{deoagn} and \citet{peng64} we find $f_{H\alpha}$ = 0.45.  \par
\noindent

\subsubsection{Cloud Geometry}

As noted previously, \citet{cheng95} have made a detailed study of
the H~I in \GH\ using the VLA with 40\arcsec\ resolution. They deduce
that the SW component is a thin rotating disk viewed nearly edge on,
with an axial ratio of about 4:1. This is based upon the shape of the
contours of the higher column density regions, but also on the shape
of the velocity displacement as a function of position as well as the
fact that such a steep inclination would explain why the peak column
density is higher than in the NE component. They also point out that
if the aspect ratio is $>$ 3 then the true (normal) surface density in
the SW component is not high enough to trigger star formation (none is
observed), contrary to the situation in the NE component. For this
proposed geometry, $A_{tot}/A_{proj} = 8$.

However, the dynamical mass for \GH\ SW calculated by Chengalur et al
(1995) within a radius of 3\arcmin\ is $7 \times 10^8$\Msun, slightly
smaller than the total HI mass of \GH\ SW.  An HI mass comparable to
the dynamical mass is of course contrary to what is now
known about the ratio of total to baryonic mass. 
%RJWREVBEG --STUART: CHANGED WORDING HERE A BIG
%RJWREVEND
This apparent
discrepancy can be reduced in several ways.  First, the HI mass refers
to the total HI mass of \GH\ SW while the dynamical mass is only
calculated for the inner 3\arcmin.  Second, the distance to \GH\ may
be as much as a factor of three closer than 20 Mpc, the distance
assumed by Chengalur et al \citep{salzer,impey}. This would reduce the HI mass
more than the dynamical mass.  
%[[[NOTE: I've asked Chengalur for his
%40" resolution HI image, in order 1) to calculate the HI mass and 2)
%use as a target for the upcoming run (I have in mind to place a FP
%ring along the major axis of GH-SW)]]] 
Nonetheless, the rotational
velocity of \GH\ SW, 13 \kms, with just a small correction for
inclination if the Chengalur et al geometry is correct, is extremely
low for the HI mass, even at the near distance---see the Tully-Fisher
type relation between rotational speed and HI mass derived by \citet{mcgaugh00}.
Indeed, a reasonable interpretation of the observed
morphology of \GH\ SW is that it has two spiral arms elongated along
the line towards \GH\ NE, with a more face-on inclination.
%*** Looking at the Chengalur et al's figures again, if I had money to bet,
%I would bet it on the two-spiral hypothesis rather than on the ``pure''
%(i.e. no nucleus) tidal feature hypothesis! ***

Accordingly, we must conclude that the geometry is uncertain.  The
extremes are represented by the 4/1 inclined disk, with
$A_{tot}/A_{proj} = 8$, and a face-on disk, with $A_{tot}/A_{proj} =
2$.  For comparison, a sphere gives $A_{tot}/A_{proj} = 4$.
An infinitely long cylinder with the long axis in the plane of the sky, gives
$A_{tot}/A_{proj} = \pi$; this factor is larger if
the cylinder is inclined to the line of sight. 
%RJWREVBEG
An ellipsoid rotated about
its major axis and with a major to minor axis ratio of 4 (perhaps a reasonable model for a tidal
feature) gives $A_{tot}/A_{proj} \approx  3.22$.  Again, this factor would
be slightly larger if the major axis is inclined somewhat to the line of sight.
In the discussion in \S4 We adopt
$A_{tot}/A_{proj} = 4$ as a fiducial factor, recognizing that there is an uncertainty of about
two in either direction. We will refer to the $A_{tot}/A_{proj}$ values of 2, 4, and 8 as
``low'', ``fiducial'', and ``high'' in the following discussion.
%RJWREVEND

\section{RESULTS AND DISCUSSION}

\subsection{Inferred limits on the Radiation Field and Comparison with Previous Results}

%RJWREVBEG --values changed throughout and some wording changes; better read whole section

Using the values for the geometrical factors just discussed, the 8 mR upper limit for
the \Ha\ flux, and values of $f_a$ and $f_{H\alpha}$ from equation
(\ref{phi_ex_calc}) we obtain values for the one sided ionizing flux
of $\Phi \, = 10.1, \, 5.04, \, \rm{and} \, 2.52 \, \times \, 10^3 \, {\rm photons\ cm^{-2} \,
s^{-1}}$ for the low, fiducial and high values of the
geometrical correction factor $A_{tot}/A_{proj}$. The first two of these
are within the range inferred by \cite{maloney}, while the
third is about a factor of two lower, though it should be remembered that
our surface brightness value is an  upper limit.  Expressing the radiation field as $J_{\nu} = J_{\nu_0}
(\nu/\nu_0)^{-\alpha}$, we obtain $J_{\nu_0} = \alpha\, \pi^{-1} \,h
\Phi$, where $h$ is Planck's constant and $\alpha$ is the
spectral index.  Assuming the SRGPF value for the spectral index of
$\alpha = 1.8$, the corresponding upper limits for the mean intensity at the
Lyman limit are $J_{\nu_0} \, = 0.96, \, 1.92, \, \rm{and} \, 3.84 \, \times \, 10^{-23} \, {\rm
ergs\ cm^{-2} \, s^{-1} \, sr^{-1}}$. The local unshielded metagalactic
photoionization rate is calculated from this as $\Gamma = 4 \pi \,
J_{\nu_0} \, \alpha_{\nu_0} \, h^{-1} \, (\alpha + 3)^{-1}$, which
yields $\log (\Gamma) \, = -13.62, \, -13.32, \, {\rm and} \, -13.02$, where $\Gamma$ has the dimensions
s$^{-1}$. Had we adopted a different spectral index (e.g., 1.0) the
corresponding numbers for our fiducial value of $A_{tot}/A_{proj} = 4$
of $\Phi$, $J_{\nu_0}$ and log$(\Gamma)$
would be $5.92 \, \times 10^3$, $1.24 \,  \times 10^{-23} $ and $-13.43$,
respectively.  If our possible detection at 4 mR should ultimately be
confirmed, the actual values would of course all be lower by a factor of
two.

A review of recent attempts to detect, or set limits on, the intensity
of the ionizing background may be found in SRGPF. Some of these are
summarized in their Table 1. It should be noted however that in the
paper by \citet{megan}, their definition of $\Phi$ is in terms of
the ionizing specific intensity, not the flux. They also assume a
value for $f_a$ of 1.0, so their
numbers in their Table 2 cannot be compared directly with ours. Their upper
limit on the \Ha\ surface brightness from \GH\ SW is about 65\,mR. The
\Ha\ surface brightness limit of VWRH (20mR) is the lowest surface
brightness limit we are aware of prior to the current work, which is
about 2.5 times higher than our new upper limit. Our use of different
assumptions about the values of $f_a$ and $f_{H\alpha}$ and 
the range of the projection factors considered here (VWRH assumed a
face--on slab) results in our current value for $\Phi$ which ranges
between about 2.5 to 10 times lower than that quoted in VWRH.

\subsection{Comparison with Estimates of AGN and Hot Star
Contributions to the Background}

A recent careful analysis of the contributions expected from both AGNs
and hot stars to the ionizing background has also been carried out in
SRGPF as well as estimation of the uncertainties in these
estimates. In principle, these contributions are directly observable
independent of cosmological parameters, but of course local H~I
prevents radiation near the Lyman limit from penetrating into the disk
of the galaxy along essentially all lines of sight from the sun.  In
practice, according to SRGPF, the two largest uncertainties in the
estimate of the contributions from the AGNs are the UV luminosity
functions for the AGNs and the assessment of the amount of attenuation
from the cumulative absorption of intervening H~I between ourselves
and the AGNs.

As noted above, SRGPF derive a best estimate for the UV spectral index
of 1.8 which we have adopted. For the value of the specific intensity
at the Lyman limit they quote

\begin{equation}
J_{\nu_0,agn} \ = \ 1.3^{+0.8}_{-0.5} \ 
\times 10^{-23} \ {\rm ergs \ cm^{-2} s^{-1} \ Hz^{-1} sr^{-1} }
\end{equation}

For the contributions from hot young stars embedded in galaxies the
luminosity function of ionizing radiation is also somewhat uncertain,
but the dominant uncertainty in this case is the fraction $f_{esc}$ of
ionizing photons which escape the galaxy. SRGPF thus characterize this
fraction as a free parameter but favor a value of order 5\%, and for
this contribution quote

\begin{equation}
J_{\nu_0,stars} \ = \ 1.1^{+1.4}_{-0.7} \ 
\times 10^{-23} \ ergs \ cm^{-2} s^{-1} \ Hz^{-1} sr^{-1} 
\end{equation}

\noindent
the sum of which is to be compared to our  2$\sigma$ upper limits of $J_{\nu_0}$
which range from  $0.96 \, {\rm to} \, 3.84 \,
\times 10^{-23} \ {\rm ergs \ cm^{-2} s^{-1} \ Hz^{-1} sr^{-1}} $.

If we examine the various combinations of the ``best'', ``upper''
 and ``lower'' 1$\sigma$ estimates in SRGPF we find that
our largest value (for the face--on projection)
is compatible for all but the largest of these combinations.
% of the two ``lower
%1$\sigma$'' estimates from the AGNs and hot stars, though it must be
%emphasized that if the large value of $A_{tot}/A_{proj}$ we have
%adopted from \citet{cheng95} is too large by a factor of two then
%several more combinations of these limits are compatible. Taking our
%result at face value however, it seems likely to us that the typical
%escape fraction is smaller than the value of 5\%\ adopted as a
%fiducial value by SRGPF. 
For our fiducial value of the projection factor,
our limit is compatible with, but slightly lower than, the sum of the
two ``best'' contributions.

%RJWREVBEG:  STUART AND SYLVAIN: I THINK THE PARAGRAPH STARTING
% WITH "In this connection, and ending with "halo objects"
% IS NOW PERHAPS MOOT, SO I HAVE DELETED IT WITH %, BUT LET ME KNOW
% WHAT YOU THINK 

%In this connection we note that for our own galaxy, model calculations
%(e.g., \citet{dsf}) predict that 3 to 20\%\ of the photons escape the
%disk of the galaxy.  Based on observations of the Magellanic stream by
%\citet{ww96}, \citet{bhm} estimate that approximately $\sim 6\%$ of
%the ionizing photons escape the galaxy.  However, it now appears that
%the original model underestimates the escape fraction required to
%explain the Magellanic Stream observations by at least a factor of
%five \citep{ben,bhput}.  Such large escape fractions are of course
%not plausible.  It seems clear that derivation of $f_{esc}$ from
%\Ha\ observations of the Magellanic stream and high velocity clouds
%has substantial uncertainties, and a smaller upper limit on $f_{esc}$
%indicated by our \GH\ observations is not in conflict with
%observations of Galactic halo objects.

\subsection{Relevance to the \Lya\ Simulations}

The range of our upper limits on the inferred z $\sim$~ 0
photoionization rate of log $\Gamma = -13.6 \, {\rm to} \, -13.0$ may be compared with that
inferred from very recent numerical simulations of the \Lya\ forest by
\citet{dave}.

The recent advances in both computing power and numerical simulation
techniques have produced results for the uncondensed gas giving rise
to the \Lyf\ which reproduce with impressive faithfulness
the observed properties of the \Lya\ absorption lines over the entire
range of observed redshifts from $\sim 0$ to $\sim 5$.  Assuming a specific
cosmological model, it is possible to infer, at any epoch, the value
of the ionizing flux which can be compared with empirical
determinations such as the one we have presented above.

The basic argument is as follows: A complete specification of the
cosmological parameters (e.g., $H_0$, $\Omega_{bary}$,
$\Omega_{\Lambda}$, the H/He ratio, and, e.g. an LCDM initial
perturbation model) completely specifies the total baryon density at
any redshift, and, in principle the course of the entire process of
gas condensation. The simulations thus yield estimates of the actual
density of hydrogen nuclei at any point in the spatial grid and at any
redshift, along with the velocity field of that material, and thus the
local column density of hydrogen nuclei. Comparing these simulations
with the observed values of the frequency of the H~I absorption lines
one can then infer the required fraction of hydrogen nuclei which are
neutral and thus the H~I photoionization rate, $\Gamma$. \citet{dave}
deduce a value of $\log \Gamma = -13.3 \pm 0.7$ at a mean redshift of
0.17, and a rough extrapolation of this estimate to zero redshift
yields a value of about $-13.5$, which is compatible with our range of upper limits.
%$\log \Gamma = -13.6$ given the uncertainties. 
Of course there are considerable uncertainties
in the simulations themselves, in addition to the basic cosmological
parameters, but these are most severe for the highest density regions
where the relatively rare higher column density H~I systems are
formed.  With further improvements in the sensitivity of the \Ha\
measurements, improved data on the density of the low redshift \Lyf\
line density, and refinements in the simulations themselves, it will
be possible to impose a constraint on the combination of the
cosmological parameters which are involved in the \citet{dave}
estimate, especially those that yield the present epoch baryon
density.

%RJWREVEND

\subsection{Caveats}

As our detailed discussion of the reduction procedures and our final
spectra make clear, there are certainly uncertainties in our fitted
value for the \Ha\ surface brightness and its upper limit. 
%RJWREVBEG
However, the uncertainty in the geometry of \GH\ SW also introduces
an additional major uncertainty.

 There is also the
possibility that our (implicit) assumption that the H~I gas has a
projected filling factor near unity is wrong.  Suppose in fact the H~I
gas in the cloud has already undergone significant fragmentation on
scales much smaller than the 21 cm resolution, with each ``cloudlet''
having H~I column densities substantially higher than that inferred
under the assumption that no such small scale structure exists. 
This implies a corresponding
reduction in the actual solid angle subtended by the high column density
fragmented gas. In such circumstances the surface brightness of \Ha\
from each little cloudlet would be only slightly increased from that
expected in the absence of fragmentation, however the total \Ha\ flux
would be reduced by the projected area filling factor, and the true
value of the photoionization value would be higher than that we have
inferred. We consider this scenario rather unlikely since one would
expect star formation to be underway in at least some portions of the
structure if such gas fragmentation had already occurred to such an
extent. %As noted by \citet{cheng95} their adopted geometry and the
%assumption of a smooth distribution of gas is consistent with the
%absence of star formation in the SW component in contrast to the NE
%component.
It is also possible that the tidal interaction between the NE and SW
components has produced filamentary or clumpy HI structure as
sometimes seen in higher linear resolution observations of tidally
interacting galaxies.

We have also considered the possibility of significant amounts of
 internal extinction of \Ha\ in the SW component by dust. Even for a
 normal dust/gas ratio, for $N(HI) = 10^{19}$ \cmsq\, the extinction
 is negligible; however, for the peak observed column density of
 $1.1\times 10^{21}$ \cmsq\ in the 40\arcsec\ resolution map of
 \citet{cheng95}, the emission in that direction could be attenuated to
 75\%\ of the extinction-free value. However, most of the 6590.4 \AA\
 ring lies toward regions of the target where the column density is
 much lower (see Figure \ref{wcce_image}), with a median column
 density of $4\times 10^{19}$ \cmsq.  In any case, a normal dust/gas
 ratio in a cloud in which no star formation seems to have occurred
 seems unlikely.
 
It would be useful to reobserve the SW component of \GH\ with improved
sensitivity in order to see if \Ha\ at the level of $\sim$4 mR is
really present. 
%RJWREVBEG
However, the ambiguity in the geometry of
\GH\ SW remains a difficuly. It would
be very interesting for numerical simulations of interacting galaxies to be run to see
if the morphology and velocity fields of the entire \GH\ complex can
be understood, and this may shed some light on the correct geometry.
Beyond the issue of the geometry however, investigation of
the initial conditions and
subsequent evolution of the rare large isolated H~I clouds such as
\GH\ in which little or no star formation has taken place would seem
a worthwhile program.
In any event, it will be important to observe with
increased sensitivity several other suitable targets, since although the SW \GH\
component has some features which make it an attractive target, the
\NII\ Galactic emission and the uncertain geometry make
a definitive determination of the background flux problematic.

%RJWREVEND

\acknowledgements

The expert assistance of Bill Kunkel and Oscar Duhalde in resolving
technical problems at the telescope is gratefully acknowledged.  RJW
thanks Romeel Dav\'{e} for helpful discussions on the \Lyf\
simulations and Don Osterbrock and Tom Slanger for enlightening
discussions and data concerning the night sky spectrum. 
SNV and SV acknowledge support from NSF
AST-9529167. Finally, we acknowledge the helpful comments of the
referee, P. Maloney, in particular for pointing out the dynamical
objection to the \cite{cheng95} geometrical model.
 
%\clearpage


\begin{thebibliography}{90}

\bibitem[Bajtlik, Duncan, \& Ostriker(1988)] {bajtlik}  Bajtlik, 
S., Duncan, R.\ C., \& Ostriker, J.\ P.\ 1988, \apj, 327, 570 

\bibitem[Barnes et al.(2001)] {barnes} Barnes, D.\ G.\ et al.\
2001, \mnras, 322, 486

\bibitem[Bechtold(1994)] {bechtold94} Bechtold, J.\ 1994, \apjs, 
91, 1 

\bibitem[Beers, Flynn, \& Gebhardt(1990)] {beers90} 
Beers, T.\ 
C., Flynn, K., \& Gebhardt, K.\ 1990, \aj, 100, 32 

\bibitem[Bertin \& Arnouts(1996)] {ber96} Bertin, E. \& Arnouts,
S. 1996, \aap, 117, 393

\bibitem[Bland-Hawthorn, Freeman, \& Quinn(1997)] {bfq} 
Bland-Hawthorn, J., Freeman, K.\ C., \& Quinn, P.\ J.\ 1997, \apj, 490, 143 

%\bibitem[Bland-Hawthorn \& Maloney(1999)] {bhm} 
%Bland-Hawthorn, J.\ \& Maloney, P.\ R.\ 1999, \apjl, 510, L33 

\bibitem[Bland-Hawthorn \& Putnam(2001)] {bhput} Bland-Hawthorn, J.\
\& Putnam, M. 2001, in ASP Conf. Series 000, Gas and Galaxy Evolution, 
ed. J.E. Hibbard, M.P. Rupen, \& J.H. van Gorkom (San Francisco:ASP), 000

\bibitem[Burton \& Hartmann(1994)] {bh94} Burton, W.\ B.\ \&
Hartmann, D.\ 1994, \apss, 217, 189

\bibitem[Carignan \& Purton(1998)] {carig98}
Carignan, C., \& Purton, C.\ 1998, \apj, 506, 125

\bibitem[Chengalur, Giovanelli \& Haynes(1995)] {cheng95} Chengalur,
J.N., Giovanelli, R., \& Haynes, M. 1995, \aj, 109, 2415

\bibitem[Dav\'{e} \& Tripp(2001)] {dave} Dav\'e, R. \& Tripp, T. M. 2001, 
\apj, 553, 000~~(in press; astro-ph/0101419)

%\bibitem[Djorgovski, G.(1990)] {djorg} Djorgovski, S.\ 1990, \aj, 99, 31

\bibitem[Donahue, Aldering \& Stocke(1995)] {megan} Donahue, M.,
Aldering, G., \& Stocke, J.\ T.\ 1995, \apjl, 450, L45

\bibitem[Dove \& Shull(1994)] {dsm94} Dove, J.\ B.\ \& Shull, 
 J.\ M.\ 1994, \apj, 423, 196 

%\bibitem[Dove, Shull, \& Ferrara(2000)] {dsf} Dove, J.\ B., 
%Shull, J.\ M., \& Ferrara, A.\ 2000, \apj, 531, 846 

\bibitem[Felten \& Bergeron(1969)] {felt} Felten, J. \ E. \& Bergeron, J. \ 1969, ApL, 4, 155


\bibitem[Giovanelli \& Haynes(1989)] {giov89} Giovanelli, R.\ 
\& Haynes, M.\ P.\ 1989, \apjl, 346, L5 

\bibitem[Giovanelli, Williams, \& Haynes(1991)] {giov91} 
Giovanelli, R., Williams, J.\ P., \& Haynes, M.\ P.\ 1991, \aj, 101, 1242 

\bibitem[Impey et al.(1990)] {impey} Impey, C., Bothun, G., Malin, 
D., \& Staveley-Smith, L.\ 1990, \apjl, 351, L33 

\bibitem[Maloney(1993)] {maloney} Maloney, P.\ 1993, \apj, 414, 41

\bibitem[McGaugh et al.(2000)] {mcgaugh00}
McGaugh, S., Schombert, J., Bothun, G., \& de Blok, W. \ 2000, \apj, 533, L99 

%\bibitem[McMahon et al.(1990)] {mcm}  McMahon, R.\ G., Irwin, 
%M.\ J., Giovanelli, R., Haynes, M.\ P., Wolfe, A.\ M., \& Hazard, C.\ 1990, 
%\apj, 359, 302 

\bibitem[Osterbrock(1989)] {deoagn} Osterbrock, D. E. 1989, 
Astrophysics of Gaseous Nebulae and Active
Galactic Nuclei (Mill Valley: University Science Books)

\bibitem[Osterbrock, Fulbright, \& Bida(1997)] {deo97} Osterbrock,
D. E., Fulbright, J.P., \& Bida, T.A. PASP 109, 614.

\bibitem[Pengelly \& Seaton(1964)] {peng64} Pengelly, R.\ M., \& 
Seaton, M.\ J. 1964, \mnras, 127, 145

\bibitem[Rand, Kulkarni, \& Hester(1990)] {rfk} Rand, R.\ 
J., Kulkarni, S.\ R., \& Hester, J.\ J.\ 1990, \apjl, 352, L1 

\bibitem[Salzer et al.(1991)] {salzer} Salzer, J.\ J., di 
Serego Alighieri, S., Matteucci, F., Giovanelli, R., \& Haynes, M.\ P.\ 
1991, \aj, 101, 1258 

\bibitem[Schlegel, Finkbeiner, \& Davis(1998)] {schlegel} Schlegel, D.\ J., 
Finkbeiner, D.\ P., \& Davis, M.\ 1998, \apj, 500, 525 

\bibitem[Scott et al.(2000)] {scott} Scott, J., Bechtold, J., Dobrzycki, 
A., \& Kulkarni, V.\ P.\ 2000, \apjs, 130, 67

\bibitem[Shull et al.(1999)] {shull} Shull, J.\ M., Roberts, 
D., Giroux, M.\ L., Penton, S.\ V., \& Fardal, M.\ A.\ 1999, \aj, 118, 1450 
(SRGPF)

\bibitem[Slanger et al.(2000)] {slangero2} Slanger, T.\ G.\ ,
Cosby, P.\ C.\ , Huestis, D.\ L.\ , \& Osterbrock. D.\ E. 2000,  \jgr, 105, 20557

\bibitem[Stocke et al.(1991)] {sto91} Stocke, J.\ T., Case, 
J., Donahue, M., Shull, J.\ M., \& Snow, T.\ P.\ 1991, \apj, 374, 72 

\bibitem[Sunyaev(1969)] {suny} Sunyaev, R. \ A. \ 1969, ApL, 3, 33


\bibitem[Tufte, Reynolds, \& Haffner(1998)] {tufte} Tufte, S.\ 
L., Reynolds, R.\ J., \& Haffner, L.\ M.\ 1998, \apj, 504, 773 

\bibitem[Turner \& MacFadyen(1997)]{turner97} Turner, N.\ J.\ 
J.\ \& MacFadyen, A.\ 1997, \mnras, 285, 125 

\bibitem[Veilleux, Cecil, \& Bland-Hawthorn(1995)] {vcbh} 
Veilleux, S., Cecil, G., \& Bland-Hawthorn, J.\ 1995, \apj, 445, 152

\bibitem[Vogel et al.(1995)] {vwrh} 
Vogel, S.\ N., Weymann, R., Rauch, M., \& Hamilton, T.\ 1995, \apj, 441, 162 
(VWRH)

\bibitem[Wakker(1990)] {wakker} Wakker, B.\ P.\ 1990, Ph.D.\ 
Thesis, University of Groningen

\bibitem[Wakker \& van Woerden(1997)] {wvw} Wakker, B.\ P.\ 
\& van Woerden, H.\ 1997, \araa, 35, 217 

\bibitem[Walsh, Staveley-Smith \& Oosterloo(1997)] {wsso}
Walsh, T., Staveley-Smith, L., \& Oosterloo, T. \ 1997, \aj, 113, 1591

\bibitem[Weiner(2001)] {benmonte} Weiner, B.\ J. 2001, in preparation

\bibitem[Weiner, Vogel, \& Williams(2000)] {ben} Weiner, B.\ J.,
Vogel, S.\ N., \& Williams, T.\ B. 2001, in ASP Conf. Series 000, Gas and Galaxy Evolution, 
ed. J.E. Hibbard, M.P. Rupen, \& J.H. van Gorkom (San Francisco:ASP), 000~~(astro-ph/0008263)


\bibitem[Weiner \& Williams(1996)] {ww96}  Weiner, B.\ J.\ \& 
Williams, T.\ B.\ 1996, \aj, 111, 1156 

%\bibitem[Wolfire(2001)] {wolfire} Wolfire, M.\ G., private communication
%to SNV

\end{thebibliography}
\end{document}